\documentclass[twocolumn,superscriptaddress,reprint,nofootinbib,prd]{revtex4-1}
\usepackage{amssymb}
\usepackage{color}
\usepackage{graphicx}
\usepackage{multirow}
\usepackage{dcolumn}
\usepackage{bm}
\usepackage[header,title,page,titletoc]{appendix}
\usepackage{amsmath}
\usepackage{bbm}
\usepackage{amsfonts}
\usepackage[bookmarks,colorlinks=false]{hyperref}
\usepackage{ulem}

\newlength{\wordlength}

\newlength{\onewordlength}

\newcommand{\ba}{\begin{eqnarray}}
\newcommand{\ea}{\end{eqnarray}}
\newcommand{\be}{\begin{equation}}
\newcommand{\ee}{\end{equation}}

\newcommand{\btheta} {{\mbox{\boldmath$\theta$}}}

\newcommand{\bn}{{\bf n}}

\newcommand{\bk} {{\mathbf k}}

\newcommand{\bp} {{\mathbf p}}

\newcommand{\bx}{{\bf x}}

\newcommand{\calO}{{\mathcal O}}
\newcommand{\calP}{{\mathcal P}}
\newcommand{\calQ}{{\mathcal Q}}
\newcommand{\calR}{{\mathcal R}}

\begin{document}

\title{A Lattice Study of $(\bar{D}_1 D^{*})^\pm$ Near-threshold Scattering}

\author{Ting Chen}
\affiliation{%
School of Physics, Peking University, Beijing 100871, China
}%

\author{Ying Chen}
\affiliation{%
Institute of High Energy Physics, Chinese Academy of Sciences, Beijing 100049, China
}%

\author{Ming Gong}
\affiliation{%
Institute of High Energy Physics, Chinese Academy of Sciences, Beijing 100049, China
}

\author{Yu-Hong Lei}
\affiliation{%
School of Physics, Peking University, Beijing 100871, China
}%

\author{Ning Li}
\affiliation{%
School of Science, Xi'an Technological University, Xi'an 710032, China
}%

\author{Chuan Liu}%
\email[Corresponding author. Email: ]{liuchuan@pku.edu.cn}
\affiliation{%
School of Physics and Center for High Energy Physics, Peking
University, Beijing 100871, China
}%
\affiliation{Collaborative Innovation Center of Quantum Matter, Beijing 100871, China}

\author{Yu-Bin Liu}
\affiliation{%
School of Physics, Nankai University, Tianjin 300071, China
}

\author{Zhao-Feng Liu}
\affiliation{%
Institute of High Energy Physics, Chinese Academy of Sciences, Beijing 100049, China
}

\author{Jian-Ping Ma}
\affiliation{%
Institute of Theoretical Physics, Chinese Academy of Sciences, Beijing 100190, China
}

\author{Zhan-Lin Wang}
\affiliation{%
School of Physics, Peking University, Beijing 100871, China
}

\author{Jian-Bo~Zhang}
\affiliation{%
Department of Physics, Zhejiang University, Hangzhou 311027, China
}

\collaboration{CLQCD Collaboration}

\begin{abstract}
In this exploratory lattice study, low-energy near threshold scattering of the $(\bar{D}_1 D^{*})^\pm$ meson system is analyzed using lattice QCD with $N_f=2$ twisted mass fermion configurations. Both s-wave ($J^P=0^-$) and p-wave ($J^P=1^+$) channels are investigated. It is found that the interaction  between the two charmed mesons is attractive near the threshold in both channels. This calculation provides some hints in the searching of resonances or bound states around the threshold of $(\bar{D}_1 D^{*})^\pm$ system.
\end{abstract}

\maketitle

\section{Introduction}

In the past decade, a series of resonances, which are called $XYZ$ particles nowadays, have been discovered by several experimental collaborations including BESIII, Belle, BaBar, CLEOc, LHCb and so on.
There are two major categories  of theses states, one of which clusters around the charmonium region of 4.0GeV
while the other is around bottomonium scale of 10.0GeV. Despite the difference in the energy region,
which is mainly caused by the mass difference of the heavy quarks ($b$ versus $c$),
there are a lot of similarities between the two categories. In fact, some of the candidates
were first found in one sector and later on also witnessed in the other.
Ever since their discoveries, $XYZ$ particles have attracted intense attention from different fields not only from experiments but also from phenomenology and lattice simulations.
While more and more of these exotic particles were confirmed in different experiments,
the nature for most of these exotic states remains obscure despite many phenomenological studies
over the years. As a non-perturbative framework, lattice QCD is supposed
 to serve as a check on these phenomenological studies and hopefully provides
 the final answer to these questions. However, due to various technical difficulties,
 mainly because of the multi-channel nature of the problem,
 a systematic lattice study remains difficult and will keep
 an active field in the near future.

Among all the $XYZ$ particles, $Z(4430)$ is in a relatively clear situation at least experimentally.
While many of the other particles still need confirmations, $Z(4430)$ is now one of the few particles that have been established with high confidence by more than one experimental collaborations.  It was first discovered by
Belle~\cite{Choi:2007wga, Mizuk:2009da} as a resonance-like structure in
the $\pi^{\pm}\psi'$ invariant mass spectrum of $B\rightarrow K\pi^{\pm}\psi'$ decays with indefinite quantum numbers.
Later on, updated results indicate that the most favored quantum numbers is $1^+$ with significance of $6.4\sigma$, while the second probable one is $0^-$ with significance $4.6\sigma$~\cite{Chilikin:2013tch}.
In 2014, LHCb observed a resonant structure in $B^0\rightarrow K^+\pi^-\psi'$decays with unambiguously determined quantum number of $1^+$ and also excluded the possibility of $\bar{D}_1 D^{*}$ threshold effect interpretation because of the positive parity~\cite{Aaij:2014jqa}. Recently, LHCb confirms their results in a model-independent way~\cite{Aaij:2015zxa}.

In contrast to the experimental situation, theoretical understanding of the state $Z(4430)$
is still far from clear. Several phenomenological investigation have been done based on the newly reported
experimental result. In Ref.~\cite{Liu:2014eka}, the authors find that the molecular candidates are more likely to decay into the radially excited states than into ground states using the quark-interchange model, thus they prefer the interpretation of $Z(4430)$ as the $\bar{D}D^*(2S)$ molecular state. However, in Ref.~\cite{He:2014nxa} the authors study the $D^*\bar{D}_1(2420)$ interactions using one-boson-exchange model and find iso-vector bound state solutions with spin parity $J^P = 1^+$. Other interpretations also exist based on different models. Since these low-energy phenomena are non-perturbative in nature, it is desirable to study this from Lattice QCD.

The motivation of this work is two-fold.
First of all, since the mass of $Z(4430)$ lies close to the threshold of $\bar{D}_1D^*$, at
least when it was first observed,  it was conjectured to be a shallow bound state of the two charmed mesons
with a quantum number of $0^-$ by various phenomenological studies, see e.g. Ref.~\cite{Liu:2008xz}.
Therefore, the first lattice study done in the quenched approximation
also focuses on this channel~\cite{Meng:2009qt}.
It was found that, in this particular channel, the interaction of the two charmed mesons is attractive
but not strong enough to form a bound state. It is legitimate to contemplate, whether this conclusion
will be changed if one uses full lattice QCD configurations instead of the quenched ones.
Admittedly, now the mass of the original $Z(4430)$ has moved up quite a bit to about $4475$MeV so it
 no longer coincides with the threshold of $D_1$ and $D^*$.
 \footnote{However, we still call the structure $Z(4430)$ instead of $Z(4475)$.}
 However, this particular threshold of
  $(\bar{D}_1 D^{*})^\pm$ remains closest to $Z(4430)$ and therefore the scattering of the $(\bar{D}_1 D^{*})^\pm$
 is still the most relevant to study. Needless to say, there are also many thresholds below this one, whose
 effects need to be taken into account in principle, however, putting in more thresholds will complicate
 the lattice computation significantly. Therefore, without any definite information
 about other more important lower channels, we focus on a single-channel lattice study in this exploratory
 work, which constitutes the second motivation of this particular study.


 According to the experimental results of Belle and LHCb,
 the most favored quantum number of $Z(4430)$ is $1^+$ instead of
 the originally proposed value of $0^-$.
 In this study, we explore both $0^-$ and $1^+$ channels.
 For $J^P = 0^-$ sector, we study the s-wave scattering of $\bar{D}_1$ and $D^{*}$ as in our former quenched study~\cite{Meng:2009qt}. This will serve as a direct comparison with the previous quenched result. For
 the $J^P = 1^+$ sector, a non-vanishing relative orbital angular momentum between the two particles is
 introduced, similar to the lattice study of $\pi\pi$ scattering
 in the $\rho$ channel~\cite{Feng:2010es, Aoki:2011yj, Pelissier:2012pi, Dudek:2012xn}.
 Our final results indicate that, the interaction between the two
 charmed mesons are attractive in both channels.

 In the $J^P=0^-$ channel, compared to the former quenched results,
 we find that the interaction between the two mesons are stronger,
 rendering the extracted scattering length negative while the quenched scattering length is still positive.
 In the $J^P = 1^+$ sector, the two mesons are also attractive.
 Hints of possible bound states have been observed in both channels.

 This paper is organized as follows. In Sec.~\ref{Sec:theoretical-framework} we briefly review the
 ingredients of L\"uscher's formalism in both periodic boundary condition and twisted boundary conditions.
 In Sec.~\ref{Sec:operators-and correlators}, single particle and two-particle operators for both $A_1$ and $T_1$ sector are analyzed. Special attention is paid for the correlation function in twisted case when using wall sources without gauge-fixing. In Sec.~\ref{simulation-details-and-results}, simulation details are given and the physical implication of the results is analyzed. Conclusions are given in Sec.~\ref{Sec:conclusion} together with some outlooks.

\section{Theoretical Framework for the Computation}
\label{Sec:theoretical-framework}

\subsection{L\"uscher's formalism for periodic boundary condition}

 Traditionally periodic boundary condition is used for all three spatial directions
 within the four-dimension Euclidean torus in lattice QCD simulations,
\begin{equation}
  \psi(\mathbf{x} + L \mathbf{e}_{i}, t) = \psi(\mathbf{x}, t),
  \label{PBC:wave-fuction}
\end{equation}
 where $\psi(\bx,t)$ designates a generic quark field.
 Then the three-momentum $\mathbf{k}$ of any degrees of freedom is
 quantized according to,
\begin{equation}
  \mathbf{k} = \left(\frac{2\pi}{L}\right) \mathbf{n}, \; (\mathbf{n} \in \mathbb{Z}^3)
  \label{PBC:quantization-condition}
\end{equation}
 In a series of publications~\cite{Luscher:1985dn, Luscher:1986pf, Luscher:1990ux, Luscher:1991cf},
 L\"uscher has proposed a general formalism to compute low-energy scattering phase shifts of two-particle systems with zero total momentum $\mathbf{P} = 0$ in a symmetric cubic box of size $L\times L\times L$.
 It relates the discrete energy eigenvalue of the two-particle system in the finite box with the elastic scattering phase of the two particles in the infinite volume. This formalism makes it possible for the numerical simulation of scattering problems from first principles of QCD. Consider two particles with mass $m_1$ and $m_2$ respectively
 in a cubic box.
 Within center-of mass frame the two particles then
 have three-momentum $\mathbf{k}_1 = -\mathbf{k}_2 = \mathbf{k}$.
 Without any interaction between the two particles,
 the total energy of the free two-particle system is simply
\begin{equation}
  E_{1+2}(\mathbf{k})=\sqrt{m^2_1+\mathbf{k}^2} + \sqrt{m^2_2+\mathbf{k}^2},
  \label{Eqn:free-total-energy}
\end{equation}
 with $\bk$ quantized as Eq.~(\ref{PBC:quantization-condition}).
 Now if we turn on the short-ranged interaction between the two particles,
 the total energy must be deviated from the free situation,
 and therefore also the three-momentum. We simply define
\begin{equation}
  E_{1\cdot 2}(\mathbf{k})=\sqrt{m^2_1+\bar{\mathbf{k}}^2} + \sqrt{m^2_2+\bar{\mathbf{k}}^2},
  \label{Eqn:interactive-total-energy}
\end{equation}
 where $E_{1\cdot 2}$ stands for the total interacting two-particle system energy
 and $\bar{\mathbf{k}}\neq \bk$ being the modified momentum.
 We can also define a variable $q^2$, the counterpart of $\mathbf{n}^2$ in the free case, as
\begin{equation}
  q^2=\bar{\mathbf{k}}^2L^2/(2\pi)^2,
  \label{Variable:q2}
\end{equation}
 which deviates from $\mathbf{n}^2$ due to the interaction.
 Within L\"uscher's formalism, there is a direct relation between the elastic scattering phase shift
 and the variable $q^2$.
 For s-wave elastic scattering, neglecting higher partial wave mixing,
 this relation reads,
\begin{equation}
  q\cot\delta_0(q)={1\over \pi^{3/2}} Z_{00}(1;q^2),
  \label{Eqn:Luescher-Formula-m00}
\end{equation}
where $Z_{00}(1;q^2)$ is the generalized zeta function, which is formally defined as
\begin{equation}
  Z_{lm}(s;q^2) = \sum_{\mathbf{n} \in \mathbb{Z}^3} \frac{\mathcal{Y}_{lm}(\mathbf{n})}{(\mathbf{n}^2 - q^2)^s}
  \label{Eqn:generalized-zeta-function-untwisted}
\end{equation}
 for $Re(s) > (l+3)/2$ and then analytically continued to the region covering $s=1$.
 Here $\mathcal{Y}_{lm}$ is a polynomial related to the spherical harmonics through
 $\mathcal{Y}_{lm}(\mathbf{n}) = |\mathbf{n}|^l \cdot Y_{lm}(\theta,\phi)$,
 $q^2$ is a real variable which can be positive or negative.
 As argued by L\"uscher~\cite{Luscher:1990ux}, efficient algorithms can be developed to evaluate this function
 for reasonable values of $q^2$. A C-package based on the algorithm given in Ref.~\cite{Beane:2011sc} can be found, for example, at~\cite{Zhanlin:2014c}.

 In numerical simulations, the two-particle energy $E_{1\cdot2}$ is obtained from suitable
 correlation functions. Several assumptions have been made to arrive at the simple  Eq.~(\ref{Eqn:Luescher-Formula-m00}). We will mention them below.
 Firstly, lattice volume should be large enough to accommodate free single-particle states which is characterized by the parameter $m_{\pi}L$ in the simulation.  Large values of $m_\pi L$ also suppresses the wrap-around contributions. Secondly, higher angular momenta mixtures have been neglected. Within $A_1$ representation of the octahedral group $O_h$,
 the next partial wave that can mix with s-wave scattering is g-wave, which is high enough to be neglected for near threshold scattering. However, one should pay special attention to subgroups of $O_h$ and/or other representations.
 This could happen when different topology or boundary conditions are applied
 which will lead to mixture of s-wave with lower partial waves, e.g. with d-wave or
 even p-wave under broken parity.
 There are another practical limitation of this formalism, which is crucial for near threshold effect of the scattering process. As Eq.~(\ref{PBC:quantization-condition}) indicates, where the smallest momentum increment is given by $|\mathbf{k}_{min}| = 2\pi/L$ which is too large for the study of near-threshold effects.
 A traditional way of circumventing this difficulty is to use twisted boundary conditions,
 which will be talked minutely in the next section.

  We mention here that extensions to the above mentioned L\"uscher formalism also exist in the literature.
  One way is to use asymmetric rectangular box of size $(\eta_1L)\times(\eta_2L)\times(L)$
  rather than a cubic box as in L\"uscher's original formalism.
  As the minimum momentum in various directions under such formalism
  can be different if $\eta_1,\eta_2 \ne 1$, the degeneracy of low-lying modes in three-dimensional momentum space can be resolved. Then one can get much more energy levels due to the breaking of the octahedral group to subgroups with lower degeneracies~\cite{Li:2003jn, Feng:2004ua}.
  If one would like to keep the simulation within a cubic box, another way to break the cubic group symmetry is
  to boost the system~\cite{Rummukainen:1995vs,Kim:2005gf,Christ:2005gi,Davoudi:2011md, Fu:2011xz, Leskovec:2012gb, Doring:2012eu, Gockeler:2012yj}, leading to the so-called moving-frame formalism. While the asymmetric box or moving
  frame formalism only increase the energy levels discretely, the twisted boundary conditions~\cite{Bedaque:2004kc, deDivitiis:2004kq} formalism can give as much momentum modes as one need, in other words giving continuous energy levels,
  which is the method we are using.

\subsection{L\"uscher's formalism for twisted boundary condition}

  Basically, instead of the usual periodic boundary conditions
  for the quark field, one uses,
\begin{equation}
  \psi_{\scriptsize \btheta}(\mathbf{x} + L \mathbf{e}_{i}, t) = e^{i\theta_i}\psi_{\scriptsize \btheta}(\mathbf{x}, t),
  \label{TBC:wave-fuction}
\end{equation}
 where the twisting angle $\btheta = (\theta_1, \theta_2, \theta_3)$ is a tunable (vector) parameter
 and $\psi_{\btheta}$ is the twisted quark field operator. Here $\btheta = (0, 0, 0)$ corresponds to the periodic boundary conditions and $\btheta = (\pi, \pi, \pi)$ corresponds to the full antiperiodic boundary conditions.
 The twisting angle $\btheta$ basically modifies the allowed three-momenta from $(2\pi/L)\bn$ to
 $(2\pi/L)(\bn+\btheta/(2\pi))$. Since $\btheta$ is freely tunable, one can get arbitrary three-momenta, in principle.

 For convenience we introduce new hatted fields as
\begin{equation}
  \hat{\psi}(\mathbf{x}, t) = e^{-i\scriptsize{\btheta}\cdot\mathbf{x}/L}\psi_{\scriptsize \btheta}(\mathbf{x}, t),
  \label{TBC2PBC:wave-fuction}
\end{equation}
 where the original fields $\psi_{\btheta}(\mathbf{x}, t)$ satisfy the twisted boundary
 condition Eq.~(\ref{TBC:wave-fuction}), while the hatted fields $\hat{\psi}(\mathbf{x}, t)$
 will satisfy the usual periodic boundary condition Eq.~(\ref{PBC:wave-fuction}).
 For Wilson-type fermions, this redefinition of the quark fields only affects the hopping
 terms in the lattice fermion actions, which amounts to a transformation of the gauge filed with
\begin{equation}
  U_{\mu}(x) \Rightarrow \hat{U}_{\mu}(x)=e^{i\theta_\mu a/L} U_{\mu}(x),
  \label{Gauge-Field-Transformation}
\end{equation}
 with $\mu=0,1,2,3$ and $\theta_\mu=(0,\btheta)$.
 Therefore, the gauge fields $U_\mu(x)\in SU(3)$ are modified by a $U(1)$ phase.
 If such gauge fields were generated according to this scenario, it will be completely equivalent to
 the twisted boundary conditions~\cite{Bedaque:2004kc, deDivitiis:2004kq}.
 This is the so-called {\it full twisting} which is a well-defined unitary approach.
 However, generating new ensembles with twisted boundary condition requires a completely new simulation
 with dynamical quarks, which is really time consuming and low efficient, as every ensemble can only have
 one specific twisting angles and new ensemble must be generated once again for different twisting angles.
 Therefore, it is desirable to use the so-called {\it partial twisting} proposal, in which (some of) the valence quarks satisfy twisted boundary conditions but the sea quarks still satisfy periodic boundary conditions. Sachrajda and Villadoro~\cite{Sachrajda:2004mi} has shown that the finite volume correction due to the partially twisted conditions is exponentially suppressed with the increasing of spacial extent $L$.
 In this scenario, it is not necessary to regenerate new gauge field configurations for different choice of the twist angle. In fact, one can simply use the configurations with periodic boundary condition. Several simulations have been done using this approach~\cite{Flynn:2005in, Kim:2010sd, Ozaki:2012ce}.
 It has also been shown recently that, {\it partial twisting} is equivalent to {\it full twisting} in some cases~\cite{Agadjanov:2013kja}.
 We simply assume that this equivalence can be carried over to our case.

 Traditional meson interpolating operators are constructed using the hatted fields in real coordinate space under periodic boundary condition as a quark bilinear
 \be
 \hat{\mathcal O}_{\Gamma}(\mathbf{x}, t)  := \bar{\hat{\psi}}_f \Gamma \hat{\psi}_{f'}(\mathbf{x}, t),\label{PBC:quark-bilinear-in-coordinate-space}
 \ee
 where $f$ and $f'$ stands for flavor indices and $\Gamma$ for specific Dirac gamma matrix matched to given quantum numbers of the operator. By summing over the spatial coordinate $\mathbf{x}$ with an extra phase of three-momentum $\mathbf{k}' = (2\pi/L)\mathbf{n}$, we can obtain the operator with specific momentum in momentum space and its relationship with the original field using the definition in Eq.~(\ref{PBC:wave-fuction}) as
\begin{eqnarray}
  \hat{\mathcal O}_{\Gamma}(\mathbf{k}', t)
  &=& \sum_{\mathbf{x}}\bar{\hat{\psi}}_f \Gamma \hat{\psi}_{f'}(\mathbf{x}, t) e^{-i\mathbf{k}'\cdot\mathbf{x}}  \nonumber \\
  &=& \sum_{\mathbf{x}}\bar{\psi}_{\scriptsize{\btheta}_f} \Gamma \psi_{\scriptsize{\btheta}_f'}(\mathbf{x}, t) e^{-i(\mathbf{k}'+(\btheta_{f'} - \btheta_{f})/L)\cdot\mathbf{x}}  \nonumber \\
  &=& {\mathcal O}_{\Gamma}(\mathbf{k}'+(\btheta_{f'} - \btheta_{f})/L, t),
  \label{PBC:quark-bilinear}
\end{eqnarray}
thus we can get the discretized momentum as
\begin{equation}
  \mathbf{k} = \frac{2\pi}{L} \left(\mathbf{n} + \frac{\btheta}{2\pi}\right), \; (\mathbf{n} \in \mathbb{Z}^3)
  \label{TBC:quantization-condition}
\end{equation}
while $\btheta = \btheta_{f'} - \btheta_{f} = (\theta_1, \theta_2, \theta_3)$ with $\theta_i$ being restricted to $0 \le \theta_i \le \pi$ without losing of generality. Following the prescriptions given in Ref.~\cite{Agadjanov:2013kja}, if we carefully select the twisting angle $\btheta_{f}$ and $\btheta_{f'}$ , we can improves our resolution in momentum space.

 One drawback with twisted boundary condition is possible partial-wave mixing due to
 the reduction of symmetry for different twisting angles,
 which makes it difficult to extract scattering parameters.
 As is well known, the irreducible representation $A_1$ of the octahedral group $O_h$ with cubic symmetry contains partial waves of $l=0,4,6,8,\cdots$ while $T_1$ contains $l=1,3,4,5,\cdots$,
 where $l$ stands for the quantum number of partial waves~\cite{Luu:2011ep}. The lowest partial wave mixing with s-wave$(l=0)$ is g-wave$(l=4)$ in the irrep $A_1$ and for irrep $T_1$, the lowest partial wave mixing with p-wave$(l=1)$ is f-wave$(l=3)$. Higher partial waves can be safely ignored since in the low-energy region, the lowest partial wave always dominates.
 When higher partial waves are neglected,  L\"uscher's formula takes its simplest form, in the s-wave as Eq.~(\ref{Eqn:Luescher-Formula-m00}) and a similar one for p-wave with $\delta_0$ replaced by $\delta_1$.
 When the higher partial waves are not neglected, we have a much more complicated form of formula, depending
 on the number of partial waves taken into account.

\begin{table*}[htb]
 \caption{Group reduction and decomposition rules of the representation for $\Gamma_s$(s-wave) and $\Gamma_p$(p-wave) based on different twisting angles $\btheta$. $A_{1g}$($A_g$) stands for trivial irrep which does not contain the $l=1$(p-wave) contribution, while $A_1$ contains both the $l=0$(s-wave) and $l=1$(p-wave) partial waves~\cite{Ozaki:2012ce}. The subscripts
 'g' stands for gerade, while 'u' for ungerade.}
 \begin{ruledtabular}
 \begin{tabular}{cccccc}
 \btheta    &$(0,0,0)$  &$(0,0,\theta)$  &$(0,0,\pi)$     &$(\pi,\pi,0)$    &$(\pi,\pi,\pi)$ \\
 Symmetry   & $O_h$     & $C_{4v}$       & $D_{4h}$       & $D_{2h}$        & $D_{3d}$  \\
 \hline
 $\Gamma_s$ & $A_{1g}$  & $A_{1}$        & $A_{1g}$       & $A_{1g}$        & $A_{1g}$  \\
 $\Gamma_p$ & $T_{1u}$  & $A_{1}\oplus E$& $A_{2u}\oplus E_u$ &$B_{1u}\oplus B_{2u}\oplus B_{3u}$ &$A_{2u}\oplus E_u$  \\
 \end{tabular}
 \end{ruledtabular}
 \label{Table:Group-reduction-formula}
 \end{table*}

 Under twisted boundary conditions, however, the cubic symmetry in reciprocal lattice space
 is broken. The symmetry reductions under different twisting angles are shown in Table~\ref{Table:Group-reduction-formula}.
 It should be noted that, for generic value of $\btheta$, say $0 < \theta_i < \pi$, the
 inversion symmetry is also lost in the momentum space and thus partial waves of different parity can mix.
 This will lead to the mixing of p-wave with s-wave phase shift even in the irrep $A_1$.
 The L\"uscher finite size formula now reads~\cite{Ozaki:2012ce, Chen:2014afa, Chen:2015jwa},
\begin{equation}
\left|
\begin{array}{cc}
q \cot\delta_0(q) - m_{00}(q) &  m_{01}(q) \cr
m_{01}(q)^{*}                 & q^3 \cot\delta_1(q) - m_{11}(q) \cr
\end{array}
\right|=0,
\label{Eqn:Luescher-Formula-Mixing-of-SP}
\end{equation}
 where $m_{00}$,$m_{01}$,$m_{11}$ are all known functions related to the generalized zeta function.
 Although methods have been put forward on how to handle such situations, as indicated
 in the above references, we would like to avoid this complication as much as possible.
 We therefore decide to take special twisting angles with $\theta_i = 0$ or $\pi$.
 This choice preserves parity and thus the even and odd partial waves will not mix.
 This amounts to setting the off-diagonal elements $m_{01}=m_{10}=0$ in the above equation.
 Hence Eq.~(\ref{Eqn:Luescher-Formula-Mixing-of-SP}) factorizes into two independent formulae,
 one for $s$-wave, the other for $p$-wave. The detailed expression for $m_{00}$ and $m_{11}$ are
 listed in Table~\ref{Table:m-function-for-SP},
 where the $w_{lm}$ can be expressed as
\begin{equation}
w_{lm}(q) = \frac{1}{ \pi^{3/2} \sqrt{2l+1} q^{l+1} } Z_{lm}^{\scriptsize{\btheta}}(1;q^{2}),
\label{Eqn:wlm-function}
\end{equation}
  and the generalized zeta function $Z_{lm}^{\btheta}(1;q^{2})$ with the twisting angle $\btheta$ is defined as
\begin{equation}
  Z_{lm}^{\scriptsize \btheta}(s;q^2) = \sum_{\mathbf{r}\in \Gamma_{\tiny \btheta}}
  \frac{\mathcal{Y}_{lm}(\mathbf{r})}{(\mathbf{r}^2 - q^2)^s}
  \label{Eqn:generalized-zeta-function-twisted}
\end{equation}
 where the lattice grids $\mathbf{r}$ in the momentum space runs over the set $\Gamma_{\scriptsize \btheta} = \{ \mathbf{r} | \mathbf{r} = \mathbf{n} + \frac{\btheta}{2\pi}, \mathbf{n} \in \mathbb{Z}^{3} \}$. As we are always keeping the system in its center of mass frame, the two scattering mesons are back to back with opposite momentum (including the twisting angles if necessary). Thus, when comparing with our formulas with those in Ref.~\cite{Briceno:2014oea} for example, the relativistic factor $\gamma$ should be set to unity.

\begin{table*}[htb]
\caption{Function $m_{lm}$ that are related to zeta function by Eq.~(\ref{Eqn:wlm-function}), different group reduction and the corresponding irreps are also listed.}
\begin{ruledtabular}
\begin{tabular}{ccccc}
\btheta    &$(0,0,0)$    &$(0,0,\pi)$     &$(\pi,\pi,0)$    &$(\pi,\pi,\pi)$ \\
Symmetry   & $O_h$     & $D_{4h}$       & $D_{2h}$        & $D_{3d}$  \\
\hline
$m_{00}$     & $A_1:q w_{00}$    &$A_1:q w_{00}$    &$A_1:q w_{00}$    &$A_1:q w_{00}$ \\
$m_{11}$     & $T_1:q^3 w_{00}$    &$E: q^3(w_{00} -w_{20})$    &$B_1:q^3(w_{00}-w_{20}-i\sqrt{6}w_{22})$  & / \\
\end{tabular}
\end{ruledtabular}
\label{Table:m-function-for-SP}
\end{table*}

\subsection{Bound states  within L\"uscher's formalism}
\label{Subsec:bound-state-formation}
 In infinite volume, a bound state of two particles can be defined as a discrete energy eigenstate of
 the Hamiltonian with energy level below the two-particle threshold.  In a finite box, however,
 all states have discrete energies such that further identification is needed.
 In a two-particle scattering process on the lattice, the interaction
 can be attractive/repulsive rendering the lowest two-particle energy level lower/higher
 than the two-particle threshold, or in terms of the variable defined in Eq.~(\ref{Variable:q2}), we have
 $q^2<0$ or $q^2>0$. To deal with the attractive
 case~\cite{Luscher:1985dn, Luscher:1990ux, Sasaki:2006jn}
 where the dimensionless momentum $q$ is pure imaginary,
 the phase shift $\delta(q)$ should be analytically continued through
 the relation $\cot\sigma(q) = i\cot\delta(q)$ and
 Eq.~(\ref{Eqn:Luescher-Formula-m00}) is modified to,
\begin{equation}
  (-iq)\cot\sigma(q)={1\over \pi^{3/2}} Z_{00}(1;q^2),
  \label{Eqn:Luescher-Formula-m00-minusq2}
\end{equation}
 where $(-iq)>0$. The phase $\sigma(q)$ for pure imaginary $q$ is physically significant since
 if there exists a true bound state at that particular energy,
 we have $\cot\sigma(q)=-1$ in the infinite volume limit.
 In a finite volume, this relation is modified to,
\begin{equation}
  \cot\sigma(q)= -1 + {6\over2\pi\sqrt{-q^2}}e^{-2\pi\sqrt{-q^2}}+\cdots,
  \label{Eqn:Bound-Finite-Volume-Correction}
\end{equation}
 the right hand side will approximate to $-1$ in the limit of $q^2\to-\infty$.
 This indicates that an infinitely negative $q^2$ (in the infinite-volume limit)
 signifies a bound state. Terms that are ignored in the above equation
 are further suppressed should the value of $\sqrt{(-q^2)}$--which is
 proportional to $(k_BL)$ with $k_B$ the binding momentum of the bound state--be large.
 Although the pole condition is fulfilled only in the infinite volume,
 we can study the finite volume corrections in a finite volume.
 Shallow bound states tend to pose a problem here because they usually have
 rather small values of $k_BL$ and therefore receive very large finite volume corrections.
 Therefore, to really identify a bound state in a lattice simulation, one
 normally needs to study the finite volume analysis using results from
 a series of volumes, see, e.g. Ref.~\cite{Sasaki:2006jn}.

\begin{figure}[htb]
   \includegraphics[scale=0.6]{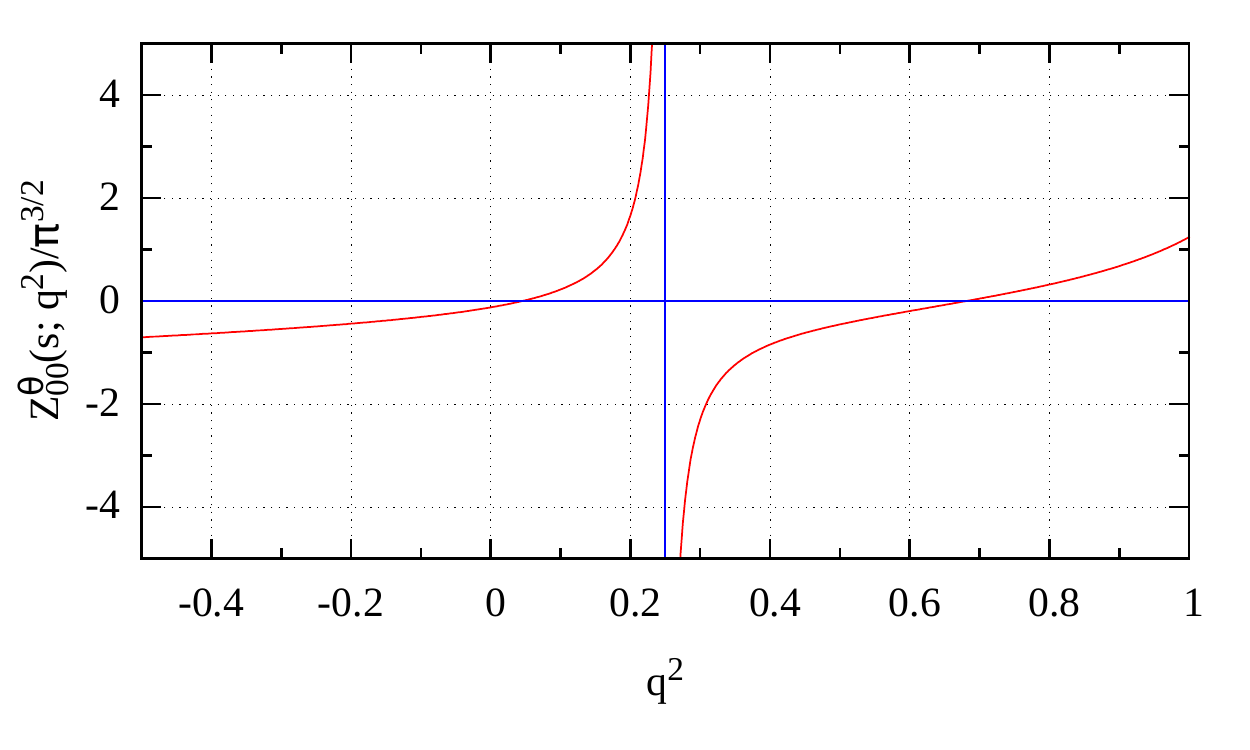}
   \caption{(color online) The function $m_{00}(q^2)$ for twisting angle $\btheta = (0,0,\pi)$. The vertical line at $q^2 = 0.25$ stands for a singular point of the zeta function under this twisting angle.}
   \label{Pic:zeta-function-range-TBC00Pi}
\end{figure}

 There is another criterion for the formation of bound state by studying
 the variation of scattering length with the energy shift of two-particle energy.
 In Ref.~\cite{Sasaki:2006jn}, the authors point out that the s-wave scattering length is positive ($a_0 > 0$)
 if the interaction between two particles is attractive but not strong enough to give rise to a bound state.
 With the increasing strength of the attraction, the sign of the scattering length turns out to be opposite($a_0 < 0$) once the bound state is formed, which can be intuitively understood by the behavior of the generalized zeta function in the region of negative parameter $q^2$, see Fig.~\ref{Pic:zeta-function-range-TBC00Pi} as an example.
 This fact provides us a distinctive identification of a loosely bound state
 even in finite volume through the observation of the lowest scattering state
 that is above the threshold, which will be discussed in the following simulation.

\section{Operators and Correlators}
\label{Sec:operators-and correlators}
  One can construct single-particle and two-particle interpolating operators
  based on the corresponding quantum numbers. Since we are interested in the interaction
  between meson $\bar{D}_1$ and $D^*$, we need single-particle operators which could create single $\bar{D}_1^0$
  and $D^{*+}$ meson from the QCD vacuum and two-particle operators for the two-particle state $(\bar{D}_1D^{*})^+$
   in various channels. Below we will first list these one-particle and two-particle operators and
   then proceed to discuss their correlation functions.

\subsection{Operators in the non-twisted case}

\subsubsection{One-particle operators}

 In lattice simulations, one should construct as much operators as possible
 to interpolate the specific particle from the QCD vacuum.
 For this preliminary study, we use the simplest quark bilinear interpolating operators
 for $\bar{D}_1^0$ and $D^{*+}$ (and also their iso-spin and anti-particle partners)
 whose quantum numbers $J^P$ are $1^+$ and $1^-$, respectively.
 In the Wilson twisted mass formulation of fermions on the lattice,
 there is some differences between the so called twisted basis and physical basis. Real computations are
 performed under the twisted basis however here we will express local interpolating fields in physical basis
 for clarity. They can be transformed into twisted basis at full twist when computing the correlation function.
 For the single charmed meson operator, we use
\begin{eqnarray}
  (\bar{D}_1^0) &:& \bar{\calP}^{(u)}_i(\textbf{x},t) = [\bar{c'} \gamma_i \gamma_5 u](\textbf{x},t), \nonumber \\
  (D^{*+})      &:& \calQ^{(d)}_i(\textbf{x},t) = [\bar{d} \gamma_i c](\textbf{x},t),
  \label{Operator:single-particle-coordinate-space}
\end{eqnarray}
 where $i=1,2,3$ indicates different spatial components, and the superscript ($u/d$)
 in parentheses stands for different light quark flavors in various charmed mesons.
 In twisted mass Wilson lattice QCD, valence quarks such as the charm quark are implemented
 using the Osterwalder-Seiler action treatment as suggested in Ref.~\cite{Frezzotti:2003ni,Blossier:2009bx}.
 The Wilson parameter of the two constituent quarks should be opposite for the quark bilinear,
 that's why we write it down in the way shown in Eq.~(\ref{Operator:single-particle-coordinate-space}).
 We can also easily get the interpolating operators for their anti-particles by applying charge conjugate, e.g. $(D_1^0):\calP^{(u)}_i(\textbf{x},t) = [\bar{u} \gamma_i \gamma_5 c'](\textbf{x},t)=[\bar{\calP}^{(u)}_i(\textbf{x},t)]^\dagger$,
 and also for isospin charged partners $\calP^{(d)}_i$ and $\bar{\calP}^{(d)}_i$ by replacing $u(\textbf{x},t)$ quark with $d(\textbf{x},t)$ quark. The same procedure can also be applied to vector charmed meson $D^{*}$.

 By a discrete Fourier transformation we can obtain single particle operators with
 definite three-momentum $\mathbf{k}$,
\begin{eqnarray}
  \bar{\calP}_i^{(u/d)}(\textbf{k},t) &=& \sum_\textbf{\scriptsize{x}} \bar{\calP}_i^{(u/d)}(\textbf{x},t)e^{-i \textbf{\scriptsize{k}} \cdot \textbf{\scriptsize{x}}}, \nonumber \\
  \calQ_i^{(u/d)}(\mathbf{k},t) &=& \sum_\textbf{\scriptsize{x}} \calQ_i^{(u/d)}(\textbf{x},t)e^{-i \textbf{\scriptsize{k}} \cdot \textbf{\scriptsize{x}}}.
  \label{Operator:single-particle-momentum-space}
\end{eqnarray}
 which is similar for the relevant charge conjugate anti-particles.
 Obviously the operators $\calP_i(\textbf{k},t)$ and $\calQ_i(\mathbf{k},t)$ form bases for
 the vector representation $T_1$ of cubic group, the lattice counterpart of $J=1$ in the continuum.
 With twisted boundary conditions implemented, the symmetry in momentum space will be further reduced from octahedral group $O_h$ to its subgroups $D_{4h},D_{2h},D_{3d}$ and so on.

\subsubsection{Two-particle operators in $A_1$ sector}

 For the two-particle system of $\bar{D}_1^0$ and $D^{*+}$ with quantum numbers
 of $1^+$ and $1^-$ respectively, we can express the two-particle system in terms of single
 particle contents with definite momentum in $A_1$ channel as
\begin{equation}
\mspace{-12mu}1^+(0^{-C})\mspace{-4mu}:\mspace{-11mu}\
\left\{
\begin{aligned}
 & \mspace{-6mu} \bar{D}_1^0 D^{*+} \!+\! \epsilon D_1^+ \bar{D}^{*0} \\
 & \mspace{-6mu} D_1^0 D^{*-} \!+\! \epsilon D_1^- D^{*0} \\
 & \mspace{-6mu} [\bar{D}_1^0 D^{*0} \!-\! D_1^- D^{*+}] + \epsilon [D_1^0 \bar{D}^{*0} \!-\! D_1^+ D^{*-}] \\
\end{aligned} \right.
\label{Operator:two-particle-contents-A1}
\end{equation}
 where $\epsilon = \pm 1$ corresponds to the charge parity of the neutral state with
 $C$-parity $C = \mp$, both of which are explored in this study.
 In our simulation positively charged partner of the iso-spin triplet is taken.
 Thus we can write down the two-particle operator as
\begin{eqnarray}
  \calO^{(A_1^-)}_\alpha(t) =\!\! &&\sum_{i,R \in G} [\bar{\calP}_i^{(u)}(R \circ \textbf{k}_\alpha,t) \calQ_i^{(d)}(-R \circ \textbf{k}_\alpha,t+1)  \nonumber \\
   &&+ \epsilon \calP_i^{(d)}(R \circ \textbf{k}_\alpha,t)\bar{\calQ}_i^{(u)}(-R \circ \textbf{k}_\alpha,t+1)]
  \label{Operator:two-particle-real-computation-A1}
\end{eqnarray}
 where the index $\alpha = 1,...,N$ with $N$ being the number of momentum modes
 considered in the simulation and the summation of $R \in G$ runs over all elements of the group
 in the question (in the case of non-twisted case, $G=O_h$).
 In our simulation for $A_1$ channel, we take $N=3$ for both non-twisted and twisted cases,
 corresponding to $\bk=(0,0,0),(1,0,0),(1,1,0)$, respectively.
 We shall call them momentum mode $0,1,2$ for simplicity.
 Note that in the above definitions, we have not included the orbital angular momentum of the two particles
 and thus only applicable to s-wave scattering  processes ($A_1$ channel).

\subsubsection{Two-particle operators in $T_1$ sector}
\label{Subsec:operator-two-particle-contents-T1}

 Similar to the study of $I=J=1$ channel $\pi\pi$
 scattering~\cite{Feng:2010es, Aoki:2011yj, Pelissier:2012pi, Dudek:2012xn} where
 vector operators are constructed from $\pi\pi$ operators,
 we can consider the $(\bar{D}_1D^*)^+$ system in the same manner.
 To be specific, we can write down the positively charged
 two-particle system similar to Eq.~(\ref{Operator:two-particle-contents-A1}) for $A_1$ sector
 with explicit single-particle contents as
\begin{widetext}
\begin{eqnarray}
1^+(1^{+C}):\bar{D}_1^0(\bk_j)D^{*+}(-\bk_j) + \epsilon D_1^+(\bk_j) \bar{D}^{*0}(-\bk_j)  - [\bar{D}_1^0(-\bk_j) D^{*+}(\bk_j) +\epsilon D_1^+(-\bk_j) \bar{D}^{*0}(\bk_j)]
\label{Operator:two-particle-contents-T1}
\end{eqnarray}
\end{widetext}
 where $j=1,2,3$ stands for the three components of the spacial momentum $\bk$ that
 forms the basis of the $T_1$ irreps;
 $\epsilon = \pm 1$ corresponds to the charge parity of the charge neutral state with $C = \mp$.
 Both cases  will be explored in this study.
 Here we only write down the positively charged part, the iso-spin partners of negatively
 charged and neutral part can be easily obtained by charge conjugate and G-parity transformations.
 From the definition in Eq.~(\ref{Operator:single-particle-momentum-space}),
 we get the operator for $T_1^+$ channel as
 \begin{widetext}
 \be
  \calO^{(T_1^+)}_{\alpha j}(t) =\sum_{i} \left\{
  \bar{\calP}_i^{(u)}\left(\bk_{\alpha j},t\right)
  \calQ_i^{(d)}\left(-\bk_{\alpha j},t+1\right)
  + \epsilon \calP_i^{(d)}\left(\bk_{\alpha j},t\right)
  \bar{\calQ}_i^{(u)}\left(-\bk_{\alpha j},t+1\right)
  - \left[\bk_{\alpha j}\Leftrightarrow -\bk_{\alpha j}\right] \right\},
  \label{Operator:two-particle-real-computation-T1}
\ee
\end{widetext}
 where $\bk_{\alpha j}= (\bk_{\alpha 1}, \bk_{\alpha 2}, \bk_{\alpha 3})$
 stands for three spatial directions of $\bk_\alpha$
 forming the basis for $T_1$ for different momentum mode $\alpha$.
 Here we only take momentum mode $1$ and $2$, i.e. $\alpha = 1, 2$,
 corresponding to momentum mode $(0,0,1)$ and $(1,1,0)$
 while momentum mode $0$, i.e. $(0,0,0)$ is automatically excluded.

\subsection{Operators in the twisted case}
\label{Subsec:twisted-operator}

 We choose to apply twisted boundary condition to the light quarks($u$ or $d$)
 while keeping the charm quark untwisted~\cite{Agadjanov:2013kja}. This avoids the
 quark-antiquark annihilation in the scattering process.
 Based on Eq.~(\ref{Operator:single-particle-coordinate-space}),
 the single particle operators are chosen to be,
 \begin{eqnarray}
  (\bar{D}_1^0) &:& \bar{\hat{\calP}}^{(u)}_i(\textbf{x},t) = [\bar{\hat{c'}} \gamma_i \gamma_5 \hat{u}](\textbf{x},t), \nonumber \\
  (D^{*+})      &:& \hat{\calQ}^{(d)}_i(\textbf{x},t) = [\bar{\hat{d}} \gamma_i \hat{c}](\textbf{x},t),
  \label{TBC-Operator:single-particle-coordinate-space}
 \end{eqnarray}
 where all the hatted fields with periodic boundary condition are related to
 the twisted fields via Eq.~(\ref{TBC2PBC:wave-fuction}).
 The same procedure can be applied to the two-particle operators in both $A_1^-$ and $T_1^+$
 channel directly.

 Compared with the case of periodic boundary conditions, there are two modifications,
 both arising from the fact that the cubic group $O_h$ is reduced to
 one of its subgroups as indicated in Table~\ref{Table:Group-reduction-formula}.

 One modification is due to the change of operator basis.
 For the operator basis in $A_1$ irrep, it remains invariant under twisted boundary condition. However,
 for the $T_1$ irrep of both single-particle and two-particle system,
 different reduction of the subgroup leads to different operator basis.
 Take the vector meson operator in Eq.~(\ref{TBC-Operator:single-particle-coordinate-space}), for example,
 for twisting angle $\btheta = (0, 0, \pi)$,
 the original operator triplet ($\calQ'^{(u/d)}_1, \calQ'^{(u/d)}_2, \calQ'^{(u/d)}_3$) should be
 decomposed into a singlet $\calQ'^{(u/d)}_3$ and a doublet ($\calQ'^{(u/d)}_1, \calQ'^{(u/d)}_2$),
 forming the basis for $A_2$ and $E$ irreps respectively.
 Special attention should be paid to the $T_1^+$ irrep of two-particle system.
 In this case the three basis of the operators are formed using different directions of the relative momentum
 for the two particles as shown in Eq.~(\ref{Operator:two-particle-real-computation-T1}).
 As the fractional momentum in twisted boundary condition
 can acquire additional spatial momentum with $(2\pi/L)\bn$,
 we choose to select those spatial momentum modes that are perpendicular
 to the fractional momentum induced by the twisting angle.
 This will keep us in the center of mass frame for the selected irrep.
 For example, for subgroup $D_{4h}$ of twisting angle $\btheta = (0, 0, \pi)$,
 we will only select ($k_{\alpha1},k_{\alpha2}$) as
 in Eq.~(\ref{Operator:two-particle-real-computation-T1}) to form the $E$ irrep for real simulation, neglecting the $A_1$ representation, and for irrep $D_{2h}$ of twisting angle $\btheta = (\pi, \pi, 0)$, we select $k_{\alpha3}$ as in Eq.~(\ref{Operator:two-particle-real-computation-T1}) to form the $B_1$ irrep for real simulation, neglecting the $B_2$ and $B_3$ representation.

 Another modification is the changing of sets of momenta used in the mode average method for $A_1$ sector. For different twisting angle, the group $G$ as in Eq.~(\ref{Operator:two-particle-real-computation-A1}) can be reduced to one of its subgroups $D_{4h}$, $D_{2h}$ or $D_{3d}$, and different momenta sets invariant under the relevant group transformation will be taken for real simulation.
 For example, we will take six momentum species, $(0, 0, \pm1), (0, \pm1, 0), (\pm1, 0, 0)$ under cubic group $O_h$ for mode average of momentum mode 1. When it comes to subgroup $D_{4h}$, as we take the twist angle $\btheta = (0, 0, \pi)$ along z-axis in this case, only $(0, \pm1, 0), (\pm1, 0, 0)$ are taken for the momentum mode averaging.
 Similar considerations also apply to the case of $D_{2h}$ and $D_{3d}$.

\subsection{Correlation functions}

 After constructing the operators for all cases, we can simply write down correlation functions with periodic boundary condition in the usual way. For the vector charmed meson $D^{*+}$, we get
\begin{eqnarray}
C^{\calQ}(\textbf{k},t)   &=& \sum_{i=1}^3 \left<\calQ_i^{(d)}(\mathbf{k},t) \calQ_i^{(d)\dagger} (\mathbf{k},0)\right>\nonumber \\
 &=& \sum_{i=1}^3 \sum_{\mathbf{x,y}} \left<\bar{d} \Gamma_i c(\mathbf{y},t) \bar{c} \Gamma_i^{\dagger} d(\mathbf{x},0) e^{-i\mathbf{k}\cdot(\mathbf{y}-\mathbf{x})} \right> \nonumber \\
 &=& - \sum_{i=1}^3  \sum_{\mathbf{x,y}} \bigg{<}\overset{(d)}M^{-1}_{\scriptsize{(\mathbf{x},0),(\mathbf{y},t)}}(\Gamma_i) \overset{(c)}M^{-1}_{(\mathbf{y},t),(\mathbf{x},0)}  \nonumber \\
  &&  \cdot (\Gamma_i)e^{-i\mathbf{k}\cdot(\mathbf{y}-\mathbf{x})}\bigg{>}
\label{Corrfunc:two-point-correlation-function}
\end{eqnarray}
 where color and spin indices are suppressed. Spatial index $i$ is summed in
 order to enhance the signal. Quantities like
\begin{equation}
\overset{(d)}M^{-1}_{(\mathbf{x},0),(\mathbf{y},t)} = \left< d(\mathbf{x},0) \bar{d}(\mathbf{y},t) \right>
\label{Prop:Single-propagator}
\end{equation}
 are quark propagators on the lattice. It should be noted from the last line of Eq.~(\ref{Corrfunc:two-point-correlation-function}) that the
 summation over all spatial points at the source is rather expensive from a computational point of view.
 We use the traditional wall-source method to reformulate it.
 To be specific, one rewrites the summation in $(\mathbf{x},\mathbf{y})$ into a summation in $(\mathbf{x},\mathbf{y},\mathbf{x'})$, and replace one of the index $\mathbf{x}$ in the two propagators in Eq.~(\ref{Corrfunc:two-point-correlation-function}) by $\mathbf{x'}$.
 Using $SU(3)$ gauge symmetry, the extra unwanted terms are gauge dependent and vanish
 after gauge field averaging. After this modification, the $D^{*+}$ two-point correlation function
 can be finally expressed as
\begin{eqnarray}
\mspace{-30mu} C^{\calQ}(\textbf{k},t) =\!\!&\!\!&\!\!\!\! \sum_{i=1}^3 \sum_{\mathbf{y}} \Bigg{<} \left( \sum_{\mathbf{x'}} \overset{(u)}M^{-1}_{(\mathbf{y},t),(\mathbf{x'},0)} \right)^* (\gamma_5\Gamma_i)  \nonumber \\
\!\!\!\!&\cdot&\!\!  \left( \sum_{\mathbf{x}} \overset{(c)}M^{-1}_{(\mathbf{y},t),(\mathbf{x},0)} e^{i\mathbf{k}\cdot\mathbf{x}}\right)(\gamma_5\Gamma_i) e^{-i\mathbf{k}\cdot\mathbf{y}}\Bigg{>},
\label{Corrfunc:two-point-correlation-function-wallsource}
\end{eqnarray}
 where we have used the so called $\gamma_5$-hermiticity for the d quark and transformed to the twisted basis such that $\Gamma_i = \gamma_5\gamma_i$ for the vector meson. This only cost one inversion for the light quark with
 zero three-momentum and one for the charm quark with momentum $\bk$. One could go further by
 averaging over different $\bk$'s which will only cost extra inversions for the charm quark
 but not the light quarks.

 The procedure discussed above is effective in the traditional non-twisted case.
 If we utilize the twisted boundary conditions in the computation, special attention should be paid
 since all field operators are changed to the hatted fields as discussed in Subsec.~(\ref{Subsec:twisted-operator}).
 One should keep in mind that it is these contractions with the hatted fields that are really computed in the simulations, in particular, using the hatted gauge fields as backgrounds.
 However, the hatted gauge fields do not live in $SU(3)$ anymore. They have extra $U(1)$ phases as shown in Eq.~(\ref{Gauge-Field-Transformation}).
 Now it is crucial to realize that, it is the un-hatted fields (with twisted boundary condition)
 that have $SU(3)$ gauge symmetry, not the hatted fields (with periodic boundary condition).
 So when it comes to the application of the $SU(3)$ gauge averaging,
 one has to express all quantities in terms of the un-hatted ones as an intermediate step,
 and transform them back to the hatted fields in the end for the real computation.
 With the help of Eq.~(\ref{TBC2PBC:wave-fuction}),
 we can relate the Wick contractions of un-hatted fields in Eq.~(\ref{Prop:Single-propagator})
 with the hatted fields as,
\begin{equation}
\overset{(d)}M^{-1}_{(\mathbf{x},0),(\mathbf{y},t)} = \overset{(\hat{d})}M^{-1}_{(\mathbf{x},0),(\mathbf{y},t)} e^{\left[i\frac{\tiny{\btheta}_d}{L}\cdot(\mathbf{x} - \mathbf{y})\right]}.
\label{Prop:Single-propagator-transformation}
\end{equation}
 Thus the two-point function $C^{\calQ}(\textbf{k},t)$ can be rewritten as
\begin{widetext}
\begin{eqnarray}
C^{\hat{\calQ}}(\textbf{k},t)
&=& \sum_{i=1}^3 \sum_{\mathbf{x,y}} \left<\bar{\hat{d}} \Gamma_i \hat{c}(\mathbf{y},t) \bar{\hat{c}} \Gamma_i^{\dagger} \hat{d}(\mathbf{x},0)  e^{-i\mathbf{k}\cdot(\mathbf{y}-\mathbf{x})}\right> \nonumber \\
&=& -\sum_{i=1}^3 \sum_{\mathbf{x,y}} \left<\overset{(d)}M^{-1}_{(\mathbf{x},0),(\mathbf{y},t)} (\Gamma_i) \overset{(c)}M^{-1}_{(\mathbf{y},t),(\mathbf{x},0)} (\Gamma_i)\cdot \exp\left[ -i\left(\mathbf{k}+\frac{{\btheta}_{c}-{\btheta}_{d}}{L}\right) \cdot(\mathbf{y}-\mathbf{x})\right]\right>, \nonumber \\
&=& -\sum_{i=1}^3 \sum_{\mathbf{x,y}} \left<\left[ \left(\sum_{\mathbf{x'}}\overset{(d)}M^{-1}_{(\mathbf{x'},0),(\mathbf{y},t)}\right) \Gamma_i \overset{(c)}M^{-1}_{(\mathbf{y},t),(\mathbf{x},0)} \Gamma_i \right]\cdot \exp\left[ -i\left(\mathbf{k}+\frac{{\btheta}_{c}-{\btheta}_{d}}{L}\right) \cdot(\mathbf{y}-\mathbf{x}) \right]\right>,
\label{Corrfunc:two-point-correlation-function-twisted}
\end{eqnarray}
\end{widetext}
 After reusing Eq.~(\ref{Prop:Single-propagator-transformation}) and $\gamma_5$-hermiticity,
 we obtain the final form for the two-point correlation,
\begin{widetext}
\begin{eqnarray}
C^{\hat{\calQ}}(R \circ \textbf{k},t)  &=& \sum_{i=1}^3 \sum_{R,\mathbf{y}} \Bigg{<} \left(\sum_{\mathbf{x'}}\overset{(\hat{u})}M^{-1}_{(\mathbf{y},t),(\mathbf{x'},0)}\exp\left[-i\frac{{\btheta}_{u}}{L} \cdot \mathbf{x'}\right] \right)^* (\gamma_5\Gamma_i)  \nonumber \\
  &&  \cdot \left(\overset{(\hat{c})}M^{-1}_{(\mathbf{y},t),(\mathbf{x},0)} \exp\left[ i\left(R \circ \mathbf{k}-\frac{{\btheta}_{u}}{L}\right) \cdot \mathbf{x}\right] \right) (\gamma_5\Gamma_i)  \cdot e^{-iR \circ \mathbf{k}\cdot\mathbf{y}} \Bigg{>},
\end{eqnarray}
\end{widetext}
 where we have incorporated the mode average operator $R$ here which belongs to the little groups reduced from the octahedral group $O_h$ depending on different twisting angle. We select the same twisting angle for the light quarks such that $\btheta_{u} = \btheta_{d}$, while the charm quark remains untwisted, thus $\btheta_{c} = (0, 0, 0)$.

 For the four-point functions, we construct a Hermitian correlation matrix
\begin{eqnarray}
  C_{\alpha\beta}(t) = \langle \calO_{\alpha}(t) \calO^{\dagger}_{\beta}(0) \rangle,
  \label{Eqn:correlation-matrix}
\end{eqnarray}
 where $\calO_{\alpha/\beta}$ represents the two-particle operators defined in Eq.~(\ref{Operator:two-particle-real-computation-A1}) and Eq.~(\ref{Operator:two-particle-real-computation-T1}),
 for $A_1$ and $T_1$ sectors respectively. Similar correlation matrix can also be constructed for the twisted case
 as discussed for the two point correlation function.
 Then the traditional procedure of Generalized Eigen-Value Problem (GEVP) can be
 applied to extract the two particle energies. Details would be discussed in  Subsec.~(\ref{Subsec:extraction-two-paiticle-energy}).

\begin{figure}[htb]
   \includegraphics[scale=0.17]{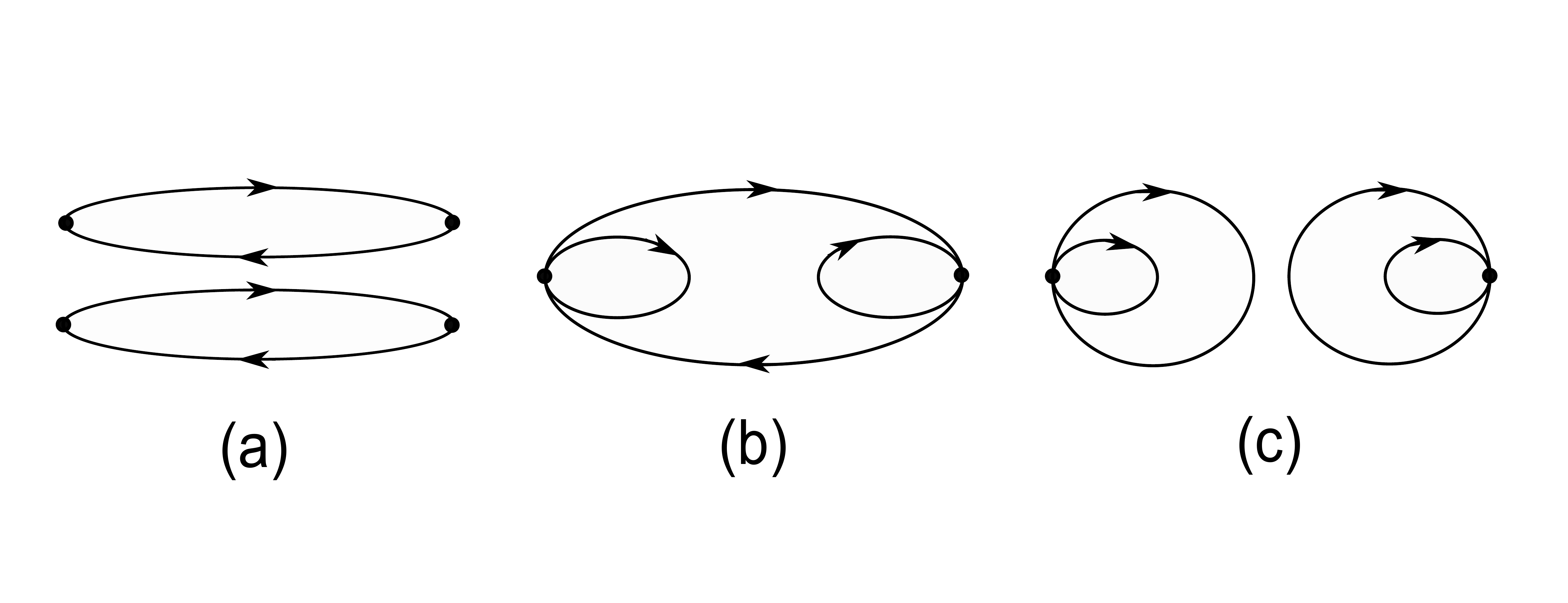}
   \caption{Typical quark flow diagrams for the four-point function:
   (a) connected, (b) singly disconnected, and (c) doubly disconnected diagrams.}
   \label{Pic:quark-flow-diagram}
\end{figure}

 The correlation matrix defined in~Eq.(\ref{Eqn:correlation-matrix}) can be expressed in terms
 of quark propagators, or contractions, using Wick's theorem. Typical quark contractions,
 also known as quark flow diagrams, are illustrated in  Fig.~\ref{Pic:quark-flow-diagram}.
 These are termed (a) connected,
 (b) single disconnected, and (c) doubly disconnected diagrams\cite{Agadjanov:2013kja,Lang:2012sv}.
 As we are studying isospin $I=1$ channel, doubly disconnected diagrams simply do not occur.
 Phenomenologically speaking, the singly disconnected diagram (b) corresponds to an exchange of charmonium state between
 the two scattering charmed mesons. The amplitude of this process is easily estimated to be small
 for close to threshold scattering. Therefore, we will simply omit the singly disconnected diagram.
 Our way of doing so is to introduce a second type of charm quark $c'$, with the same mass but
 opposite Wilson parameter as that of $c$, using the Osterwalder-Seiler type action.
 This has the extra advantage of automatic $O(a)$ improvement within the twisted mass formulation
 once we tune to the maximal twist~\cite{Frezzotti:2003ni}.

\section{Simulation details and results}
\label{simulation-details-and-results}

In this study, we utilize $N_f=2$ twisted mass gauge field configurations generated by European Twisted Mass Collaboration (ETMC) at $\beta=4.05$ for three different pion mass values. Details of the relevant parameters are summarized in Table~\ref{Table:Configuration-infomation}.

\begin{table}[h]
\centering
\caption{Simulation parameters in this study. All configurations used are of the size $32^3\times 64$ with lattice spacing $a\simeq 0.067$fm (or $\beta=4.05$). The statistics for three ensemble are all 200.}
\begin{tabular}{cccc}
\hline
\hline
    \#\rm{Ensemble}   &$\qquad\rm{I}\qquad$     &$\qquad\rm{II}\qquad$     &$\qquad\rm{III}\qquad$  \\
\hline
    $a\mu$            &$0.003$       &$0.006$        &$0.008$                  \\
    $m_\pi$[MeV]      &307.0         &423.6          &488.4                    \\
    $m_\pi L$         &3.31          &4.57           &5.27                     \\
\hline
\hline
\end{tabular}
\label{Table:Configuration-infomation}
\end{table}
 Maximally twisted Wilson quarks are used by setting the bare quark mass term in the action to its critical value,
 rendering the physical observables automatically $O(a)$ improved in the continuum limit.
 For the valence charm quark, we have used the Osterwalder-Seiler like action~\cite{Frezzotti:2003ni}.
 The up and down quark masses are fixed to the values of the sea-quark values while
 that for the charm quark is fixed using the experimental mass of the spin-averaged value
 of $J/\psi$ and $\eta_c$ on the lattice, i.e. ${3 \over 4}m_{J/\psi} + {1 \over 4}m{\eta_c}$.

\subsection{Charmed meson mass and dispersion relations}
\label{Subsec:single-particle-properties}

\begin{figure}[htb]
   \includegraphics[scale=0.6]{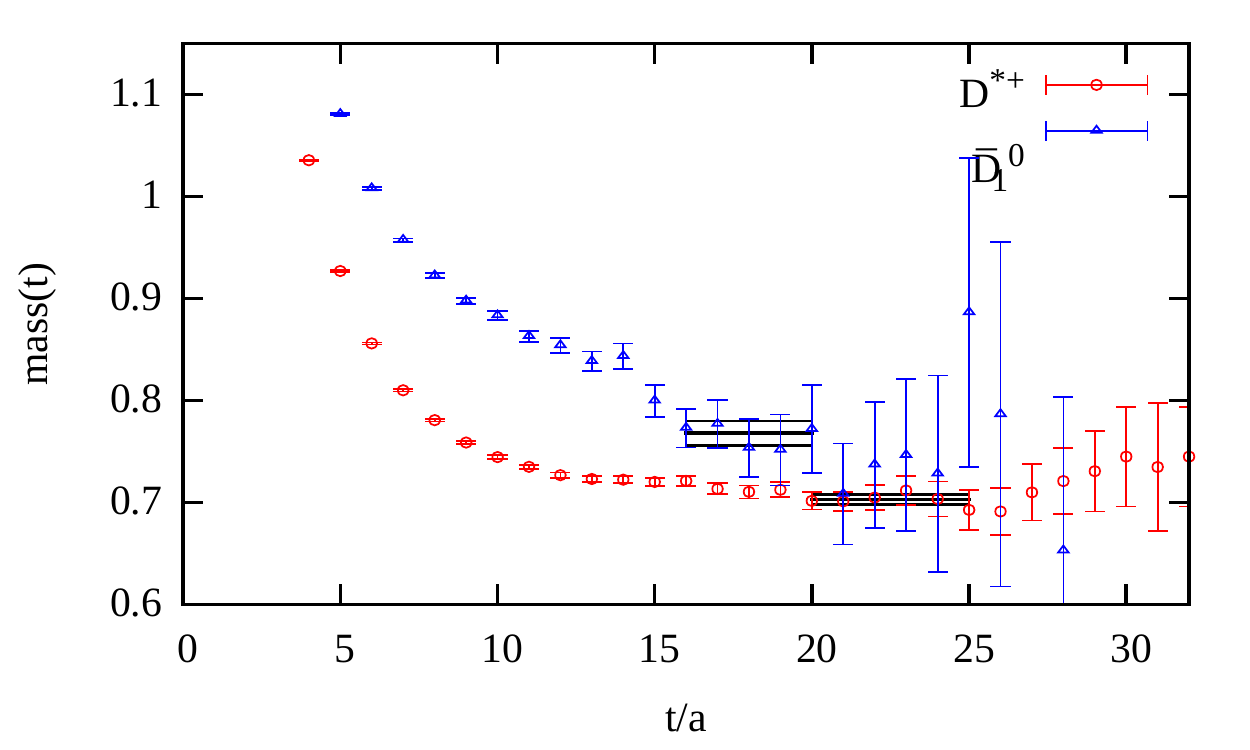}
   \includegraphics[scale=0.6]{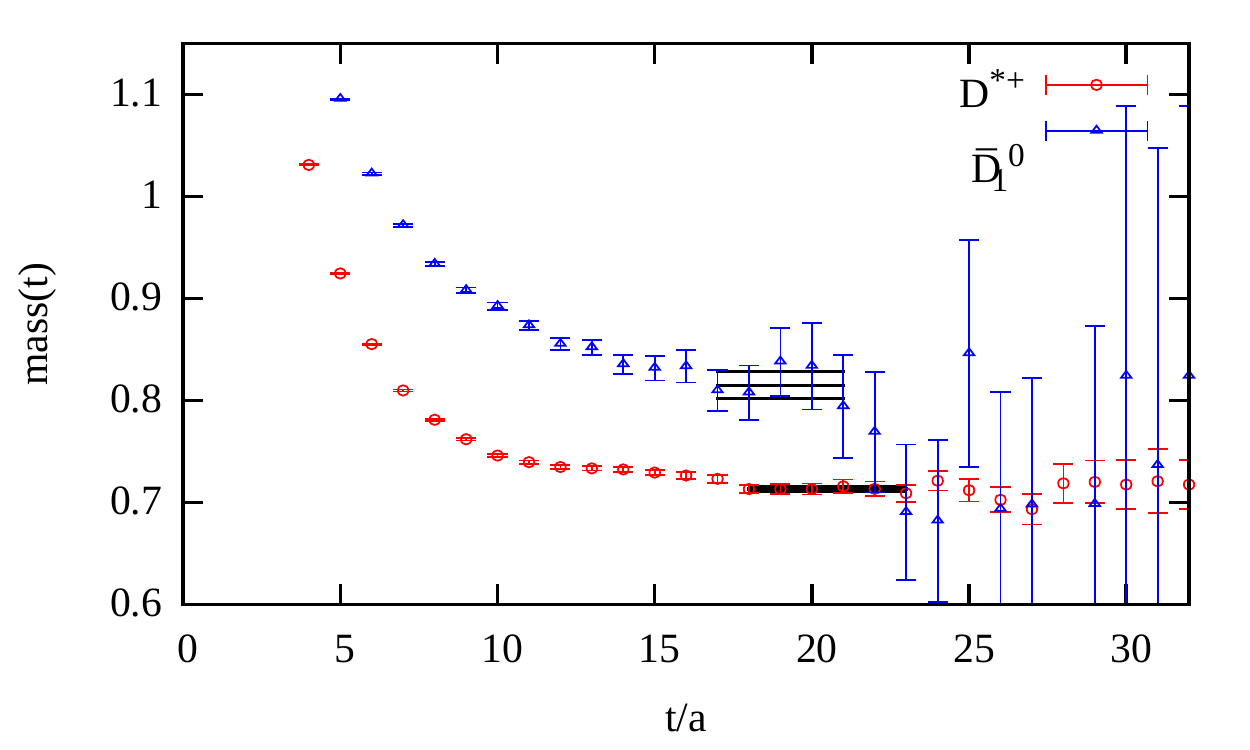}
   \includegraphics[scale=0.6]{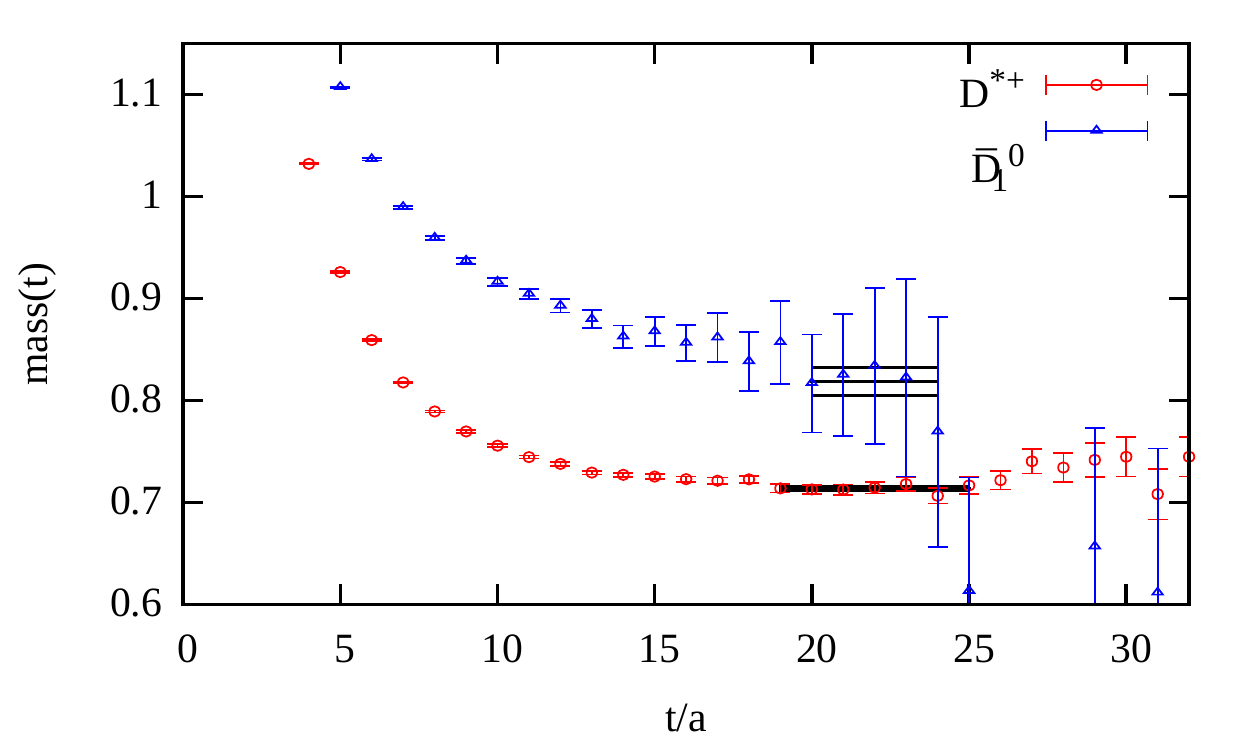}
   \caption{(color online) Mass plateaus for $\bar{D}_1^0$(triangle) and $D^{*+}$(circle), from top to bottom for Ensemble I, II, III. The central values and errors are shown by black line segments. Mass values are shown in lattice unit.}
   \label{Pic:Two-point-mass-plateau}
\end{figure}

 We have calculated the one-particle correlation functions for $\bar{D}_1^0$ and $D^{*+}$ as defined in Eq.~(\ref{Corrfunc:two-point-correlation-function}), for a series of definite three-momentum $\textbf{k}$ and twisting angle $\btheta$. After inserting a complete set of states, any single-particle correlation
 function can be written in the following form (assuming infinite temporal size)£º
\begin{eqnarray}
  C(t)=\langle \calO(t) \calO^\dagger(0) \rangle = \sum_n C_n e^{-E_n t},
\end{eqnarray}
 where $E_n$ stands for the one-particle spectrum, $n=0$ is for the ground state, i.e. particle mass for non-twisted case. For finite temporal extension, we can extract the ground states by defining the effective mass
 \begin{eqnarray}
  m_{\rm{eff}}(t) = \cosh^{-1} \left( \frac{C(t-1)+C(t+1)}{2C(t)} \right),
  \label{Plateau:direct-method-for-mass}
 \end{eqnarray}
 which in the large $t$ limit is dominated by a constant that can be regarded as the mass of the meson.
 We have also checked the logarithmic effective mass defined as $m_{\rm{eff}}(t) = \ln\frac{C(t)}{C(t+1)}$
 and found that these two methods yield compatible results while the former one is more robust
 especially in cases where plateau sets in at large $t$.
 Three time-slices for setting the source(with statistics of $3*200$) are taken and the corresponding results are
 averaged for a better signal.
  The plateau behavior for the masses of $\bar{D}_1^0$ and $D^{*+}$ are illustrated in Fig.~\ref{Pic:Two-point-mass-plateau} for three of our ensembles.

\begin{figure}[htb]
   \includegraphics[scale=0.6]{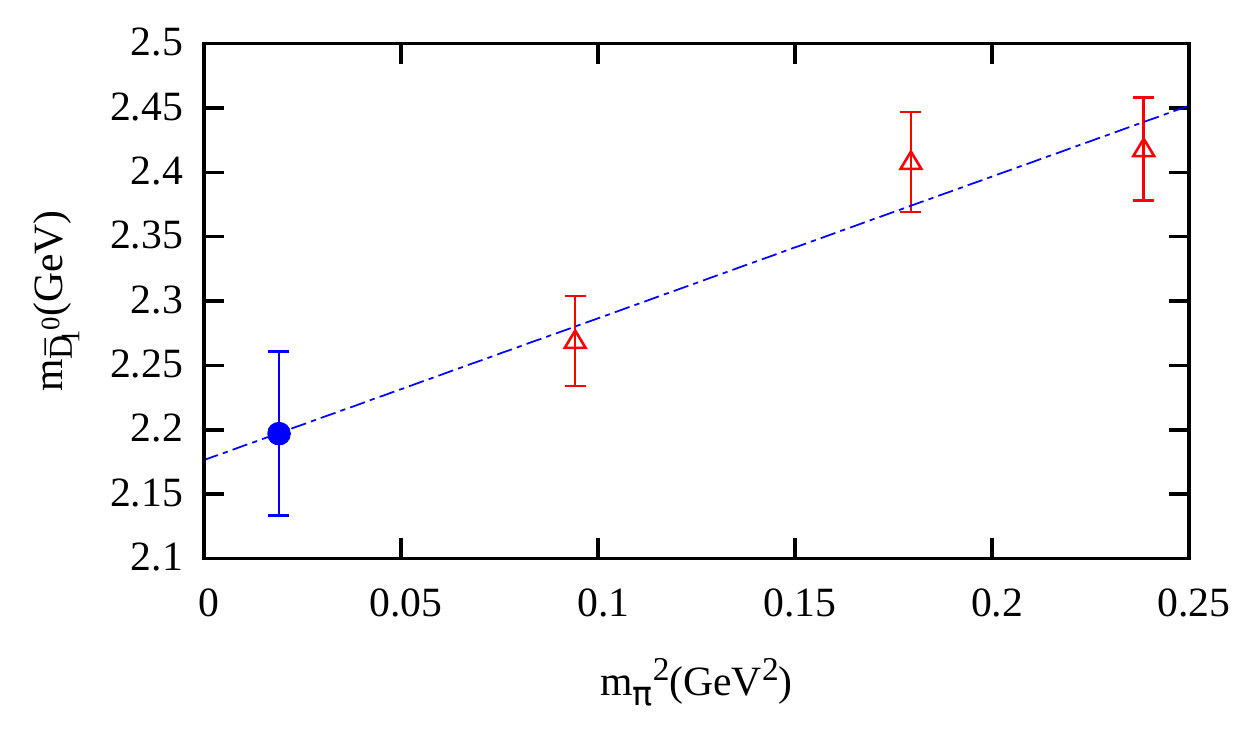}
   \includegraphics[scale=0.6]{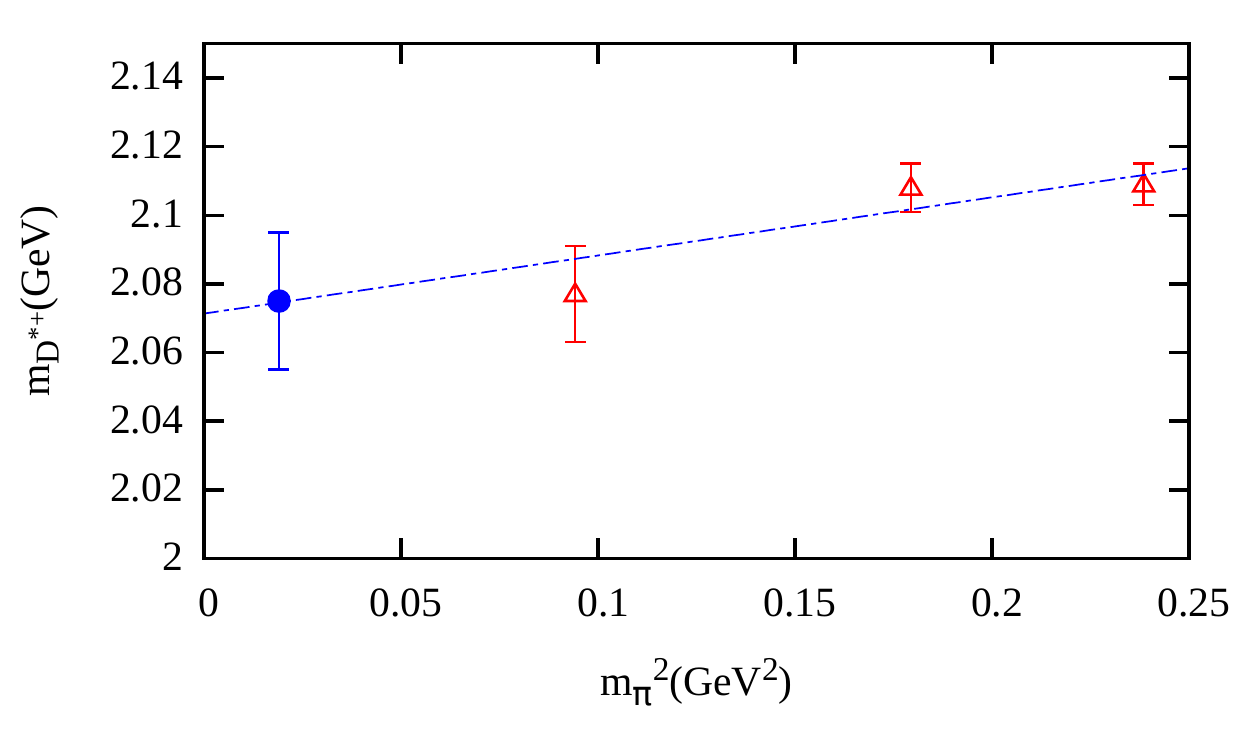}
   \caption{(color online) Chiral extrapolation for $\bar{D}_1^0$(upper panel) and $D^{*+}$ (lower panel) for three ensembles.}
   \label{Pic:Two-point-chiral-extrapolation}
\end{figure}

 After obtaining the mesons' mass from the three ensembles,
 chiral extrapolations are carried out for $m_{\bar{D}_1^0}$ and $m_{D^{*+}}$
 with linear function in $m_{\pi}^2$ to the physical pion mass, as is illustrated in Fig.~{\ref{Pic:Two-point-chiral-extrapolation}}. The extrapolated result is reasonable
 though the errors are still large, especially for the $\bar{D}_1^0$.
 The mass for the two mesons in the physical point are extrapolated to be
\begin{eqnarray}
  m_{\bar{D}_1^0} &=& 2.197 \pm 0.064 ~\rm{GeV}, \nonumber \\
  m_{D^{*+}}      &=& 2.075 \pm 0.020 ~\rm{GeV}.
\end{eqnarray}
 The mass of $D^{*+}$ comes out to be compatible with its physical value
 while that for $\bar{D}_1^0$ is lower than the narrower axial vector resonance $D_1^0(2420)$.
 Note also that the errors for the $m_{\bar{D}_1^0}$ are much larger than those
 for $m_{D^{*+}}$. This is due to the noisy nature of the $D_1$ correlator.
 One would need a more sophisticated operator basis, see e.g. Refs.~\cite{Mohler:2012na, Kalinowski:2015bwa},
 in order to reduce the noise.

\begin{figure}[htb]
   \includegraphics[scale=0.6]{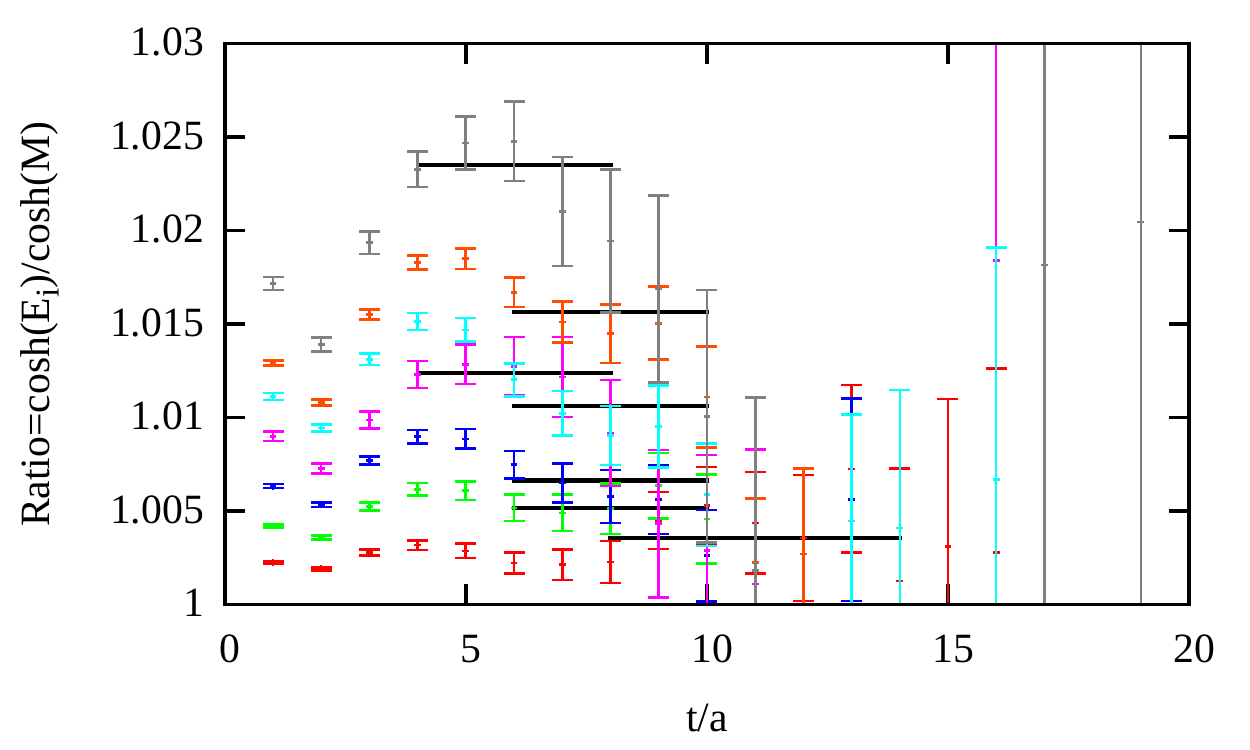}
   \includegraphics[scale=0.6]{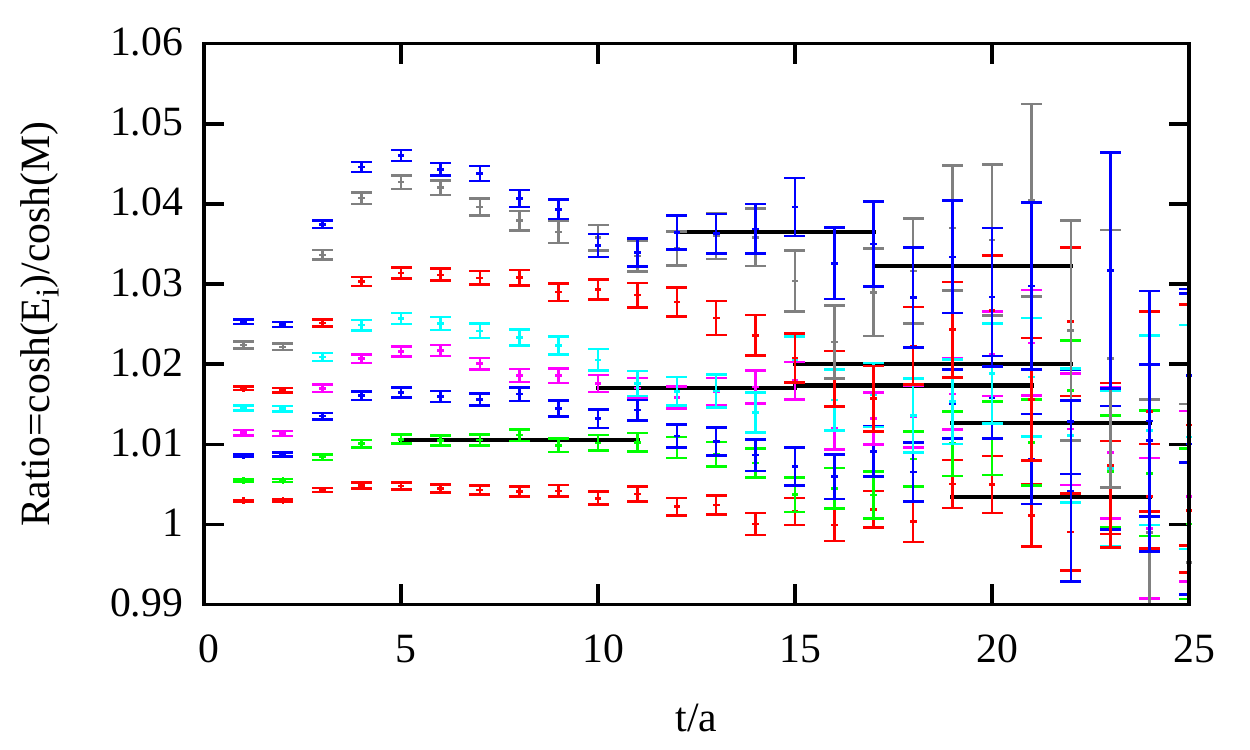}
   \caption{(color online) Ratio plateaus for $\bar{D}_1^0$(upper panel) and $D^{*+}$ (lower panel) for Ensemble II. The ratio plateaus extracted here gives the energy levels information shown in the middle panel of Fig.~\ref{Pic:Two-point-dispersion-relation}}.
   \label{Pic:Two-point-ratio-plateau}
\end{figure}

 When it comes to the correlation function $C(\mathbf{k}, t)$ with non-zero three-momentum,
 both for non-twisted and twisted case, we define a ratio $\calR(\mathbf{k}, t)$
 of correlation function with momentum $\mathbf{k}$
 to the one with zero three-momentum that gives the particle mass information as discussed above,
 \begin{eqnarray}
  \calR(\mathbf{k}, t) = \frac{[C(\mathbf{k},t-1)+C(\mathbf{k},t+1) ]\cdot C(\mathbf{0},t)}{[C(\mathbf{0},t-1)+C(\mathbf{0},t+1) ]\cdot C(\mathbf{k},t)}.
  \label{Ratio:two-point-correlation-function}
 \end{eqnarray}
 In this way, noise from the fluctuation of the ground state will be partially cancelled.
 We found this particularly useful for the noisier axial-vector meson $\bar{D}_1^0$.
 In Fig.~\ref{Pic:Two-point-ratio-plateau}, plateaus for $\calR(\mathbf{k}, t)$
 for Ensemble II are shown. The situations for other ensembles are similar.
 After extracting the plateau information for $\calR(\mathbf{k}, t)$,
 the energy at different three-momentum $\mathbf{k}$ can be obtained from
 \begin{eqnarray}
  E(\mathbf{k}) = \cosh^{-1} \left[ \calR(\mathbf{k}) \cdot \cosh(m) \right],
 \end{eqnarray}
 where the errors of $E(\mathbf{k})$ are computed from those of $\calR(\mathbf{k})$ and $m$.

 With the energy levels obtained for both normal and twisted case,
 we can further study the dispersion relations for these mesons,
 using either the discrete dispersion relation
\begin{equation}
4\sinh^2 {E_{\mathbf{k}} \over 2} = 4\sinh^2 {m \over 2} + \rm{Z_{lat.}} \cdot \sum_{i=1}^{3} 4\sin^2 {k_i \over 2},
\label{Dispersion:discrete-relation}
\end{equation}
or the continuum version
\begin{equation}
E_{\mathbf{k}}^2 = m^2 + \rm{Z_{con.}} \cdot {\textbf{k}}^2,
\label{Dispersion:continuum-relation}
\end{equation}
 where $\rm{Z^{1/2}_{lat.}}$ and $\rm{Z^{1/2}_{con.}}$ are the corresponding effective speed of light
 parameters.

\begin{figure}[htb]
   \includegraphics[scale=0.6]{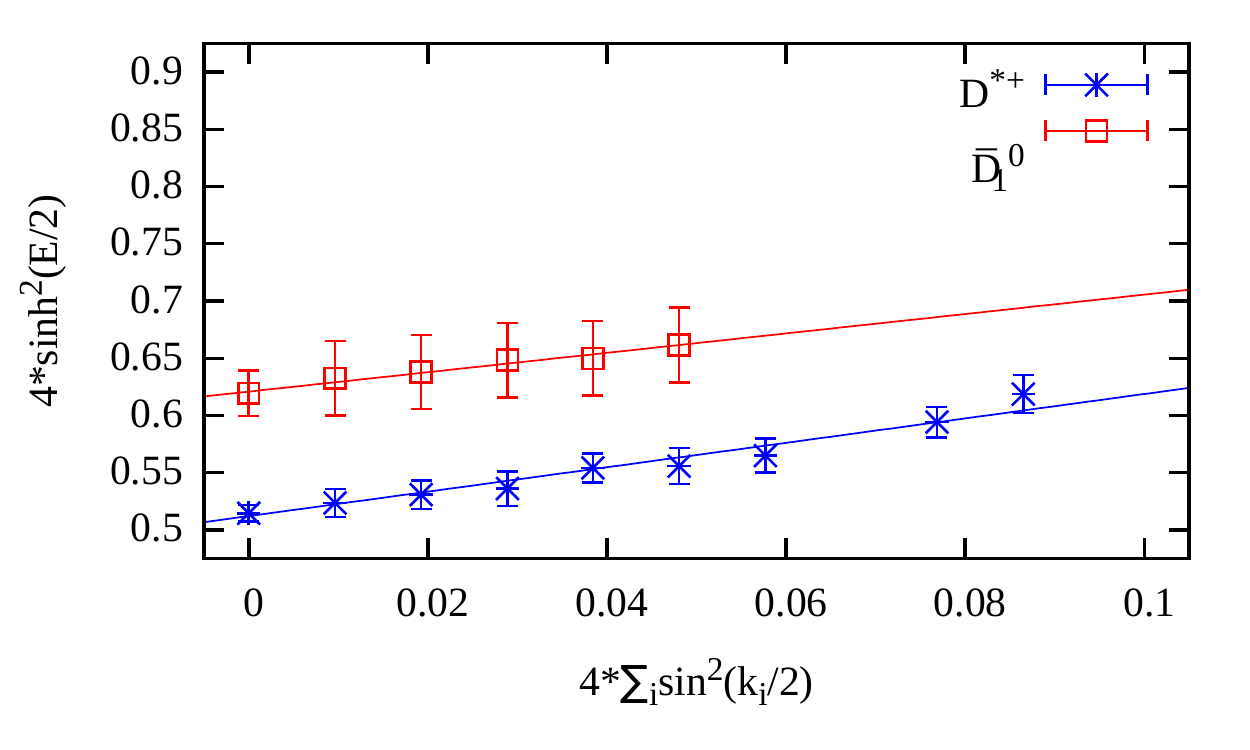}
   \includegraphics[scale=0.6]{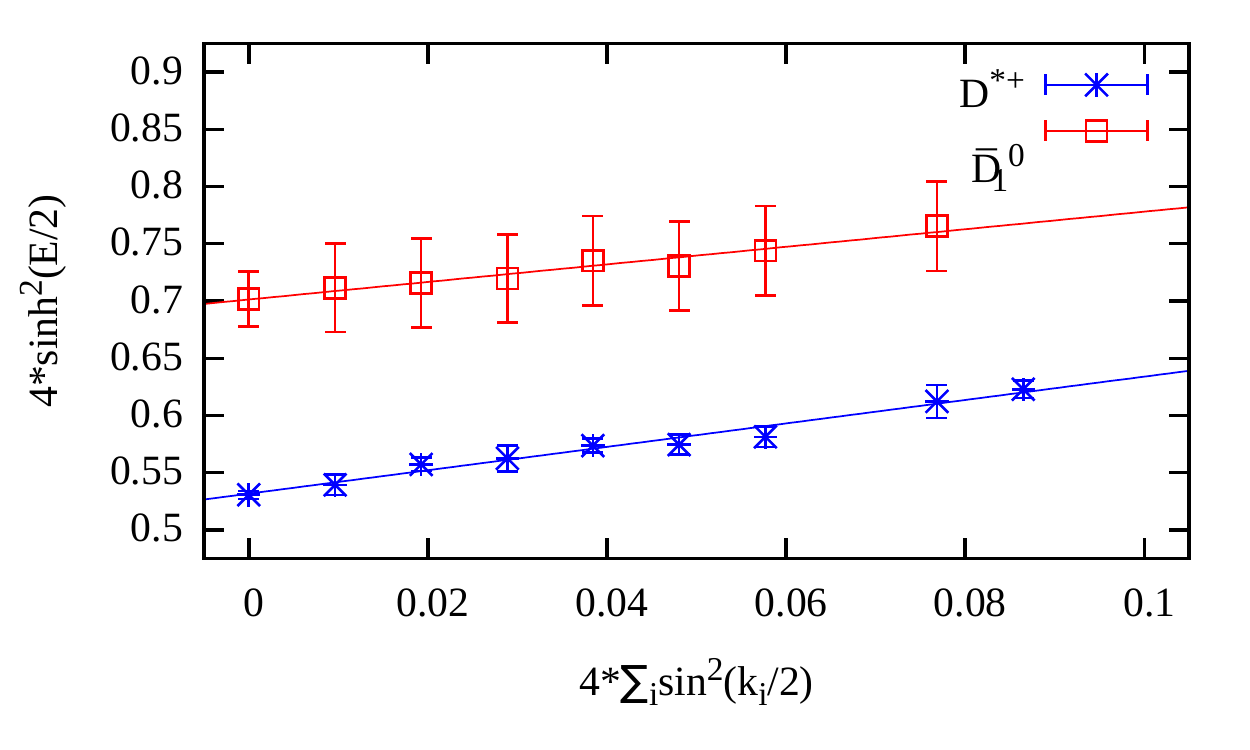}
   \includegraphics[scale=0.6]{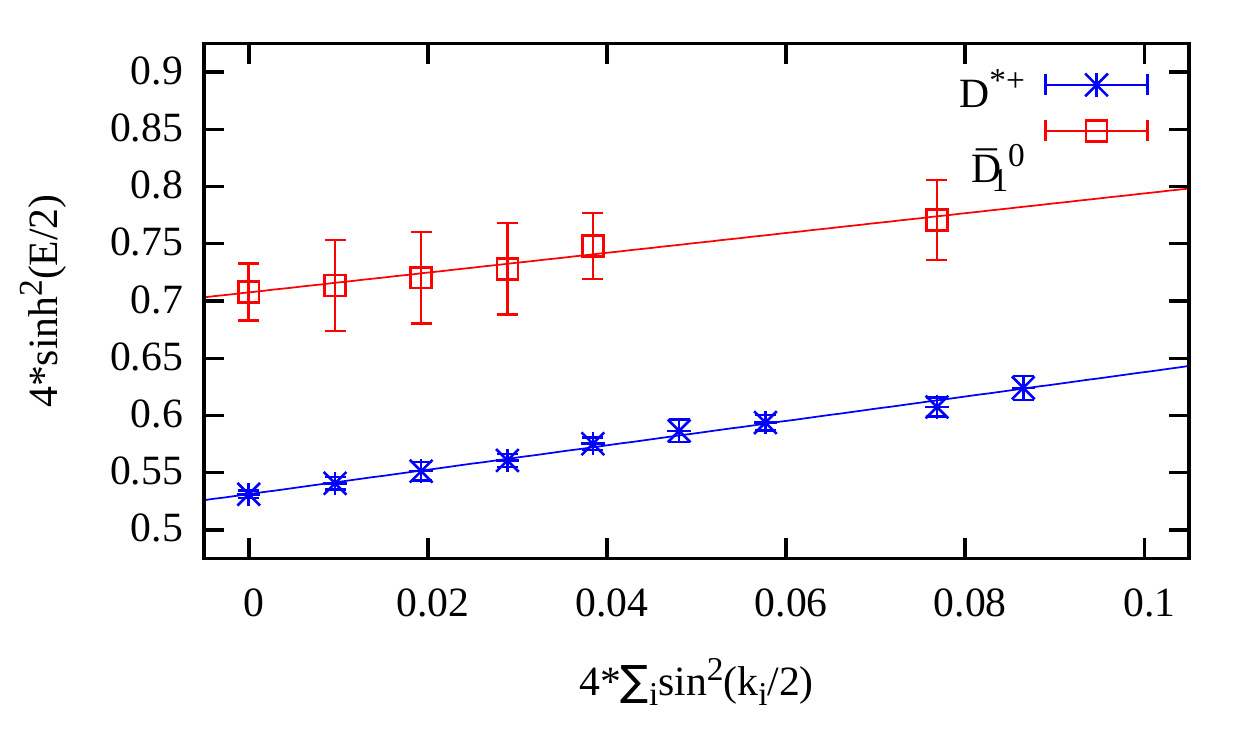}
   \caption{(color online) Discrete dispersion relation as in Eq.~(\ref{Dispersion:discrete-relation}) for the $\bar{D}_1^0$(open squares) and $D^{*+}$ (stars) meson, from top to bottom for Ensemble I, II and III. The values of the square of effective speed of light $\rm{Z_{lat.}}$ are listed in Table~\ref{Table:speed-of-light-dispersion}, comparing with the $\rm{Z_{con.}}$ fitted from Eq.~(\ref{Dispersion:continuum-relation}). }
   \label{Pic:Two-point-dispersion-relation}
\end{figure}

 The fitting results of the discrete dispersion relation based on Eq.~(\ref{Dispersion:discrete-relation})
 are shown in Fig.~\ref{Pic:Two-point-dispersion-relation} for all three ensembles.
 In this study, we have taken $\mathbf{n} = 0,1,2$ and $\btheta = (0,0,\pi),(\pi,\pi,0),(\pi,\pi,\pi)$,
 resulting in more low-momentum data points than our former results~\cite{Chen:2014afa, Chen:2015jwa}.
 Some missing points of for $\bar{D}_1^0$ in the figure are due to bad signal
 such that no stable plateau can be extracted even if the ratio method of Eq.~(\ref{Ratio:two-point-correlation-function}) is utilized.
 The fitting for data points using continuous dispersion relation based on Eq.~(\ref{Dispersion:continuum-relation}) are very similar, so we only show the comparison of the square of effective speed of light, i.e. $\rm{Z_{lat.}}$ vs. $\rm{Z_{con.}}$, for the two mesons in Table~\ref{Table:speed-of-light-dispersion}.
\begin{table}[h]
\begin{ruledtabular}
\centering
\caption{Square of effective speed of light for two mesons, with the comparison of discrete and continuous version.}
\begin{tabular}{ccccc}
    \#\rm{Ensemble}                     &                    &$\rm{I}$    &$\rm{II}$  &$\rm{III}$  \\
\hline
                                        & $\rm{Z_{lat.}}$    &0.85(67)    &0.77(49)   &0.87(53)    \\
\raisebox{1.6ex}[0pt]{$\bar{D}_1^0$}    & $\rm{Z_{con.}}$    &0.77(61)    &0.68(44)   &0.77(47)    \\
\hline
                                        & $\rm{Z_{lat.}}$    &1.07(14)    &1.02(8)    &1.07(8)    \\
\raisebox{1.6ex}[0pt]{$D^{*+}$}         & $\rm{Z_{con.}}$    &0.98(13)    &0.93(7)    &0.97(7)    \\
\end{tabular}
\label{Table:speed-of-light-dispersion}
\end{ruledtabular}
\end{table}

 For the vector meson, the results for $\rm{Z_{lat.}}$ and $\rm{Z_{con.}}$ are comparable
 and both are compatible with $1.0$ within errors.
 This indicates that the operator we used indeed interpolates a vector meson rather well.
 For the axial vector meson $\bar{D}_1^0$, however, the error of the effective speed of light
 is huge compared with that of $D^{*+}$. This is due to the bad signal of the $\bar{D}_1^0$
 meson as is seen from the Fig.~\ref{Pic:Two-point-ratio-plateau} and
 Fig.~\ref{Pic:Two-point-dispersion-relation}. 
 
 It is known that there are two $D_1$ mesons experimentally,
 the wider resonance $\bar{D}_1^0(2430)$ and the narrower one $\bar{D}_1^0(2420)$,
 which in reality couples to $D^*\pi$ two-particle states. In an earlier lattice study~\cite{Mohler:2012na}, 
 it is also found that the contamination of $D^*\pi$ two-particle states on the wider $D_1$ state 
 is substantial.  So there is a potential worry whether our $D_1$ state also has this problem.
 We look into this possibility and conclude that this is not the case due to the 
 following reasons:
 
 First of all, to suppress the contaminations from the two-particle states, 
 we have utilized the wall-source with definite momentum/twist. 
 This is known to greatly suppress the coupling to the multi-particle states.
 
 Second, unlike the situation in Ref.~\cite{Mohler:2012na} where the lowest level of $D^*\pi$ 
 state lies below the mass of $D_1$ thus the so-called level crossing of two types of states is bound to happen,
 in our case, the $D^*\pi$ two-particle states actually lie
 above the $D_1$ states by over $100$MeV as listed in Table~\ref{Table:mass-comparison}.
 Therefore possible level crossing of these two-particle states with that of $D_1$ is avoided.
 Surely the $D^*\pi$ states still have some effects on the $D_1$ single-particle state, 
 but it is not as dramatic as in Ref.~\cite{Mohler:2012na}. 
 In other words, we believe that we still acquire a single $D_1$
 state though it is rather noisy. In order to improve this situation, much more statistics
 and/or better operator basis following Refs.~\cite{Mohler:2012na, Kalinowski:2015bwa} should
 be taken.

 \begin{table}
 \caption{Mass difference comparison for three ensembles.}
 \begin{tabular}{cccc}
 \hline
 \hline
    \#\rm{Ensemble}           &$\qquad\rm{I}\qquad$     &$\qquad\rm{II}\qquad$     &$\qquad\rm{III}\qquad$  \\
 \hline
    $a\mu$                    &$0.003$       &$0.006$        &$0.008$                  \\
    $m_\pi$[MeV]              &307           &424            &488                       \\
 $m_{D^{*+}}$[MeV]            &2077(14)      &2108(7)        &2109(6)                   \\
 $m_{\bar{D}_1^0}$[MeV]       &2269(35)      &2408(39)       &2418(40)                  \\
 $(m_\pi + m_{D^{*+}}) - m_{\bar{D}_1^0}$[MeV] &115        &124           &179            \\
 \hline
 \hline
 \end{tabular}
 \label{Table:mass-comparison}
 \end{table}

\begin{figure}[htb]
   \includegraphics[scale=0.6]{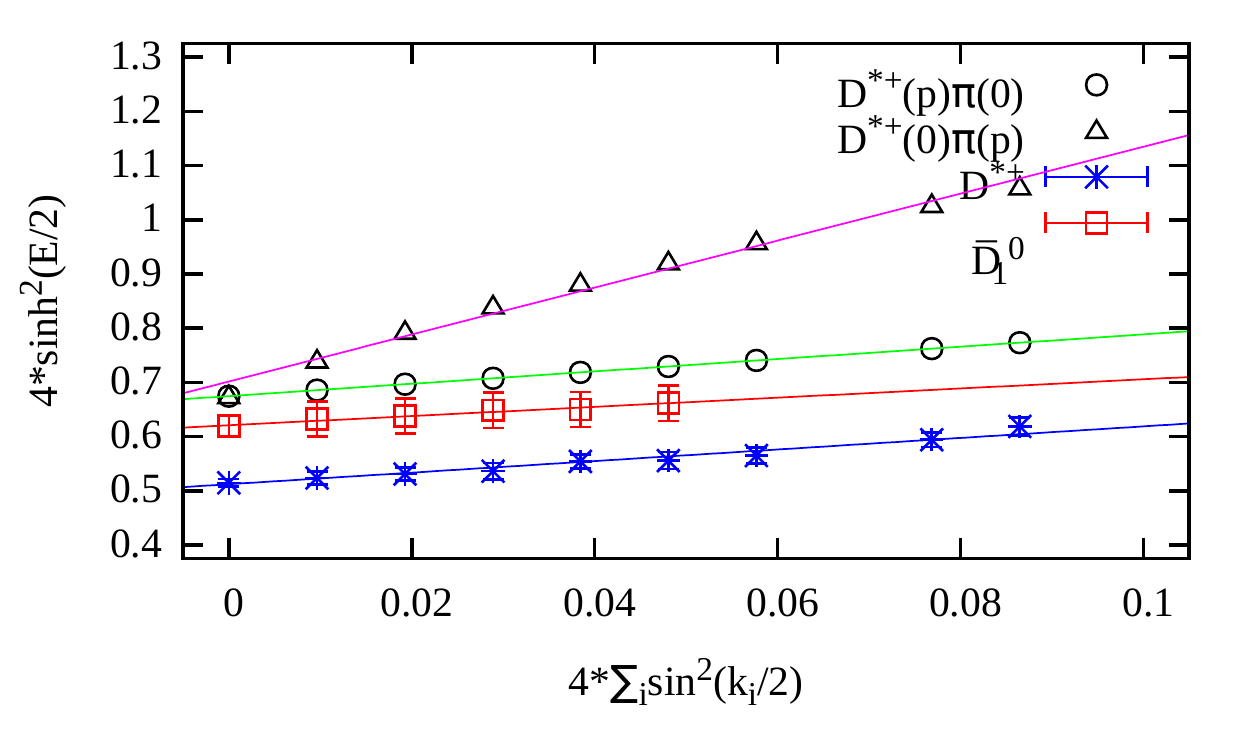}
   \caption{(color online) Two particle dispersion relation compared with the single particles' for Ensemble I.
   The energies of two-particle states form a band that is bounded below
   by the open circles.}
 \label{Pic:Dispersion-relation-compasion}
 \end{figure}

 Third, albeit its large error, the dispersion relation for $\bar{D}_1^0$ still looks like
 a single-particle one, not a two-particle one.
  To check this in further detail, we illustrate the comparison of dispersion relation
 for single particle state with that of the two particle states
 in Fig.~\ref{Pic:Dispersion-relation-compasion}. It is expected that a single particle state is definitely
 different from a two-particle state in terms of dispersion relations and this is indeed what we see.

 The energy of a two-particle state with total three-momentum $\bp$ will also depend on another
 momentum, call it $\bk$.
 Here for simplicity we will only consider two limiting cases:
 with one particle is moving with $\bp$ while the other one is at rest.
 In fact, the static $\pi$ and a moving
 $D^*$ will give us the lowest bound of these two particle states $|D^*(\bp-\bk)\pi(\bk)\rangle$,
 which is $E_{D^*}(\bp)+m_\pi$. This is shown as open circles in Fig.~\ref{Pic:Dispersion-relation-compasion}
 for Ensemble I. Taking other values of $\bk$ will modify the total energy
 of the system to $E_{D^*}(\bp-\bk)+E_\pi(\bk)$ which is even larger, depending
 on the choice of $\bk$. They form a band that is bounded below
 by the values of $E_{D^*}(\bp)+m_\pi$ (the open circles). 
 Data points for the other case of taking $\bk=\bp$ are shown
 as triangles in Fig.~\ref{Pic:Dispersion-relation-compasion}.
 It is clearly seen from the figure that our dispersion relation for $\bar{D}_1^0$ state
 (the open squares) indeed looks like
 that of a one-particle state, lying well below the two-particle bands. 
 Situations for the other two ensembles are similar.
 We therefore believe that, albeit the somewhat large error of our $D_1$ correlation functions,
 they still provide us with a reasonable one-particle state.

 Because of the large errors of the $D_1$ correlators, it might also
 hinder our search of the two-particle energy-levels of the $\bar{D}_1D^*$ system.
 Indeed, we do observe noisy behavior of the corresponding four-point functions.
 However, we managed to obtain the energy shift by constructing suitable ratios
 of the four-point functions with respect to the two-point functions. It turns
 out that large statistical fluctuations due to $D_1$ can be partly canceled by
 this ratio method which will be elaborated in the next subsection.

\subsection{Extraction of two-particle energy levels}
\label{Subsec:extraction-two-paiticle-energy}

 We adopt the usual GEVP method on correlation matrix Eq.~(\ref{Eqn:correlation-matrix})
 to extract the two-particle energy eigenvalues. In order to get more stable plateau,
 a new matrix $\Omega(t,t_0)$ is introduced,
\begin{eqnarray}
  \Omega(t,t_0)=C(t_0)^{-{1\over2}}C(t)C(t_0)^{-{1 \over 2}},
\end{eqnarray}
 where $t_0$ is the so-called reference time slice.
 Normally one picks a $t_0$ such that the signal is good and stable.
 In our simulation, a search of $t_0$ over a reasonable range is performed and
 the one that yields the smallest $\chi^2$ value in the fitting is chosen~\cite{Dudek:2007wv}.
 The energy eigenvalues for the two-particle system are then obtained by diagonalizing
 the hermitian matrix.

 The eigenvalues of the matrix exhibit the usual exponential decay behavior,
\begin{eqnarray}
  \lambda_i(t,t_0) \propto e^{-E_i(t-t_0)},
\end{eqnarray}
 from which the exact two-particle energy $E_i$ can be extracted.
 In practice, we construct the following ratio,
 \begin{eqnarray}
  \calR(t,t_0) = \frac{\lambda_i(t,t_0)}{C_{\bar{D}_1^0}(t) C_{D^{*+}}(t)} \propto e^{-\Delta E_i \cdot t},
 \end{eqnarray}
 where $C_{\bar{D}_1^0}$ and $C_{D^{*+}}$ are the corresponding one-particle correlation function
 with momentum mode 0 (ground state with zero momentum) for $A_1$ sector and momentum 1 (next lowest momentum) for $T_1$ sector. The effective energy shift $\Delta E_i$ can be extracted from the ratio
\begin{eqnarray}
  \Delta E_i(t) = \ln \frac{\calR(t)}{\calR(t+1)},
\end{eqnarray}
 where the error of $\Delta E_i$ are estimated using the conventional jackknife method in all cases,
 and thus all the errors are only statistical in the following sections.
 From the definition above, $\Delta E_i$ is the difference of the two-particle energy
 measured from the threshold of the two mesons,
\begin{eqnarray}
  \Delta E_i = E_i - m_{\bar{D}_1^0} - m_{D^{*+}}.
\end{eqnarray}
 We have also tried to look at the effective mass plateau
 from the diagonal elements of the correlation matrix in Eq.~(\ref{Eqn:correlation-matrix}).
 It turns out that the plateau is only stable for the lowest mode.
 However, if we use the ratio method discussed above, stable plateaus can be seen in almost all cases.
 We believe this is mainly due to our poor signal of the $D_1$ meson already discussed in the
 previous subsection. The ratio method have managed to cancel out some of these
 statistical fluctuations in two-point function of $D_1$ and $D^*$.

 The energy shifts together with other relevant information are summarized in Table~\ref{Table:deltaE-infomation-for-A1-channel} and Table~\ref{Table:deltaE-infomation-for-T1-channel}
 for the scalar and vector channel respectively.
 With the energy difference $\Delta E_i$ obtained, we can further define the effective momentum
 \begin{eqnarray}
  \mspace{-24mu}  \sqrt{m_{\bar{D}_1^0}^2+\bar{\mathbf{k}}^2} +\mspace{-2mu} \sqrt{m_{D^{*+}}^2+\bar{\mathbf k}^2}=\Delta E_i \mspace{-2mu}+\mspace{-2mu} m_{\bar{D}_1^0} \mspace{-2mu}+\mspace{-2mu} m_{D^{*+}},
 \end{eqnarray}
 where the $\bar{\mathbf{k}}^2\equiv(2\pi/L)^2q^2$ is effective relative momentum squared
 for the two mesons. It is this quantity that will eventually enter
 L\"uscher's formula Eq.~(\ref{Eqn:Luescher-Formula-m00}).

 For near threshold scattering,
 the effective range expansion exists for $\cot\delta(k)$,
\begin{eqnarray}
 {k^{2l+1}\cot\delta_l (k)}=a^{-1}_l+{1\over2} r_{l} k^2 +\cdots,
 \label{Eqn:effective-range-expansion-k}
\end{eqnarray}
 where $a_l$ is the scattering length and $r_l$ is the effective range for partial wave $l$.
 For convenience,  we would like to express this formula in terms of dimensionless quantity $q^2$,
\begin{eqnarray}
 {q^{2l+1}\cot\delta_l (q^2)} &=& B_l+{1\over2} R_{l} q^2 +\cdots,
 \label{Eqn:effective-range-expansion-q}
\end{eqnarray}
 with  $B_l=[L/(2\pi)]^{2l+1}a^{-1}_l$ and $R_l=[L/(2\pi)]^{2l-1}r_l$,
 which will be more convenient in our fitting process.

\subsection{Results for the Scalar channel}
\label{Subsec:scalar-channel-result}

 As in the simulation we do contractions of the propagators part by part for the terms
 shown in Eq.~(\ref{Operator:two-particle-real-computation-A1}),
 it's easy to check the charge parity by setting $\epsilon = -1 ~ \rm{or} ~ 1$.
 We find that there is no signal for the final correlation function when $\epsilon = -1$,
 corresponding positive charge parity.
 Therefore, all the following results are all for negative charge parity sector
 namely $I^G(J^{PC}) = 1^+(0^{--})$.

\subsubsection{Two-particle energy spectra}

\begin{figure}[htb]
  \centering
  \includegraphics[scale=0.6]{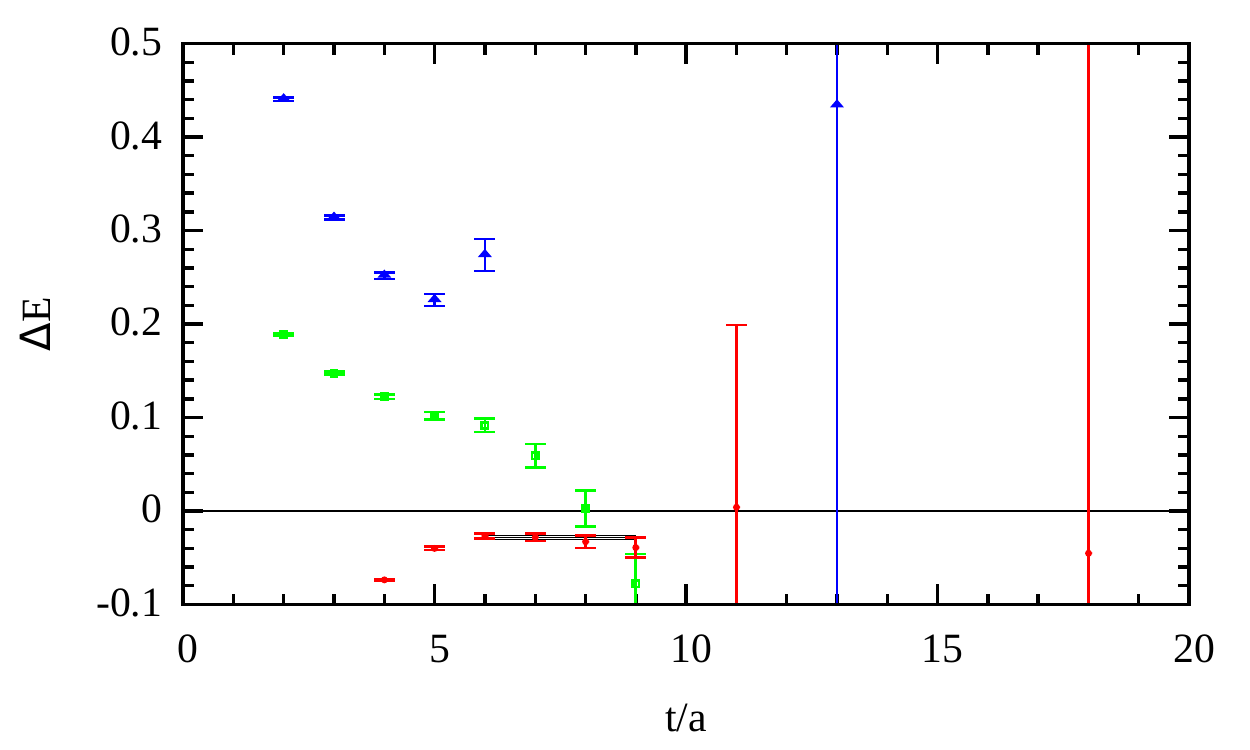}
  \includegraphics[scale=0.6]{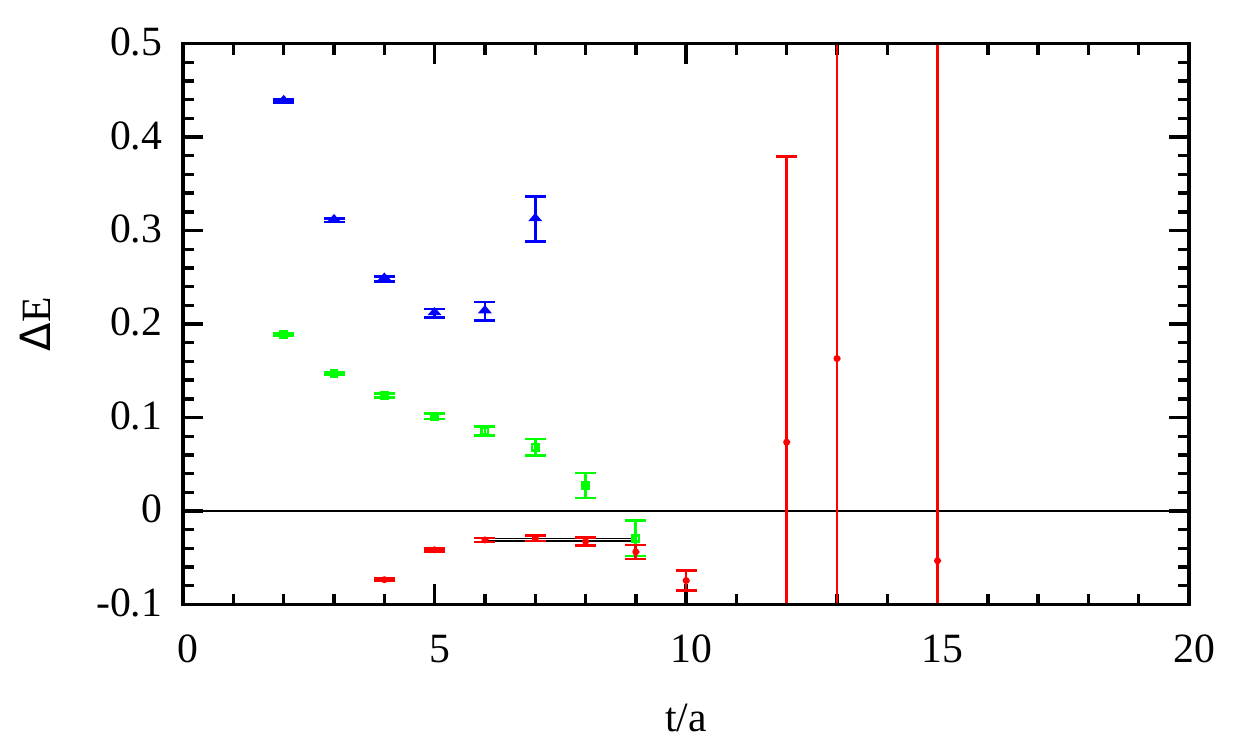}
  \includegraphics[scale=0.6]{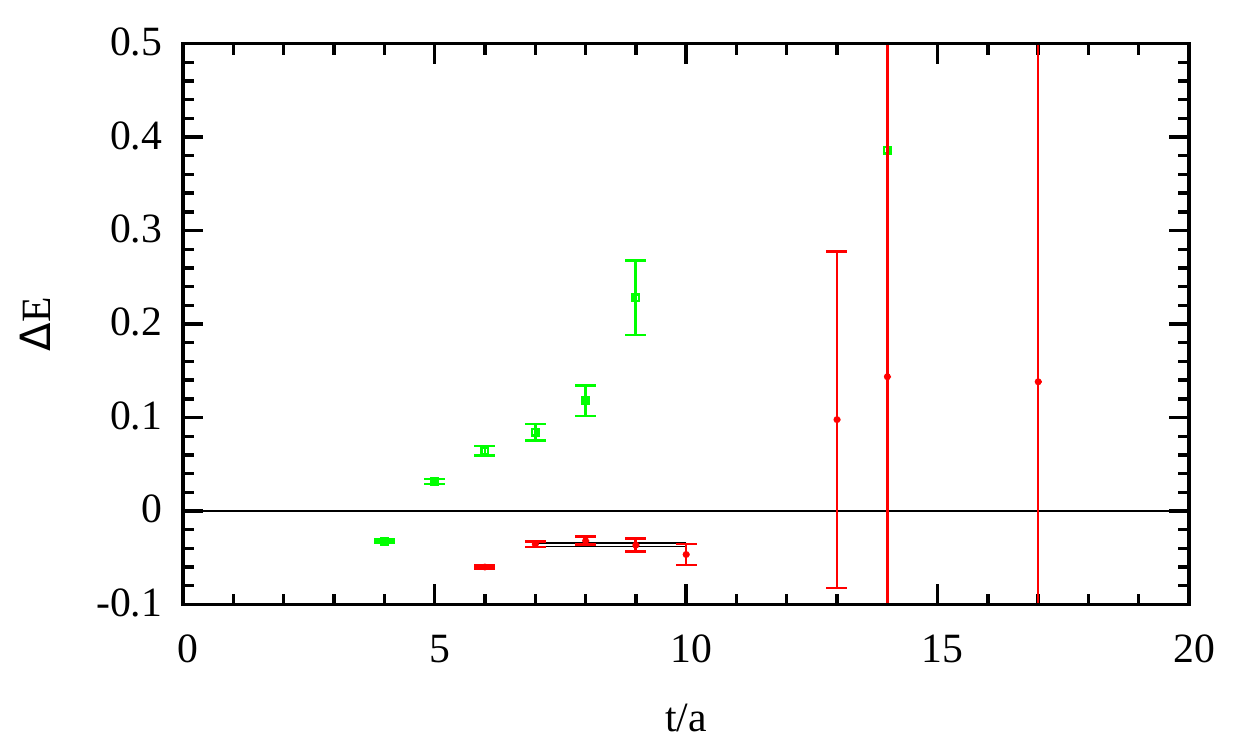}
  \caption{(color online) Effective mass plots for the energy shift $\Delta E_\alpha$ for $\btheta = \mathbf{0}$ for three ensembles, from top to bottom for Ensemble I, II and III. For Ensemble III, only the $2\times2$ sub-matrix is solved because of the numerical stability. Grey horizontal bars are the final results for the ground states.}
  \label{Pic:Energy-plateau-of-A1-channel}
\end{figure}

\begin{figure}[htb]
  \centering
  \includegraphics[scale=0.6]{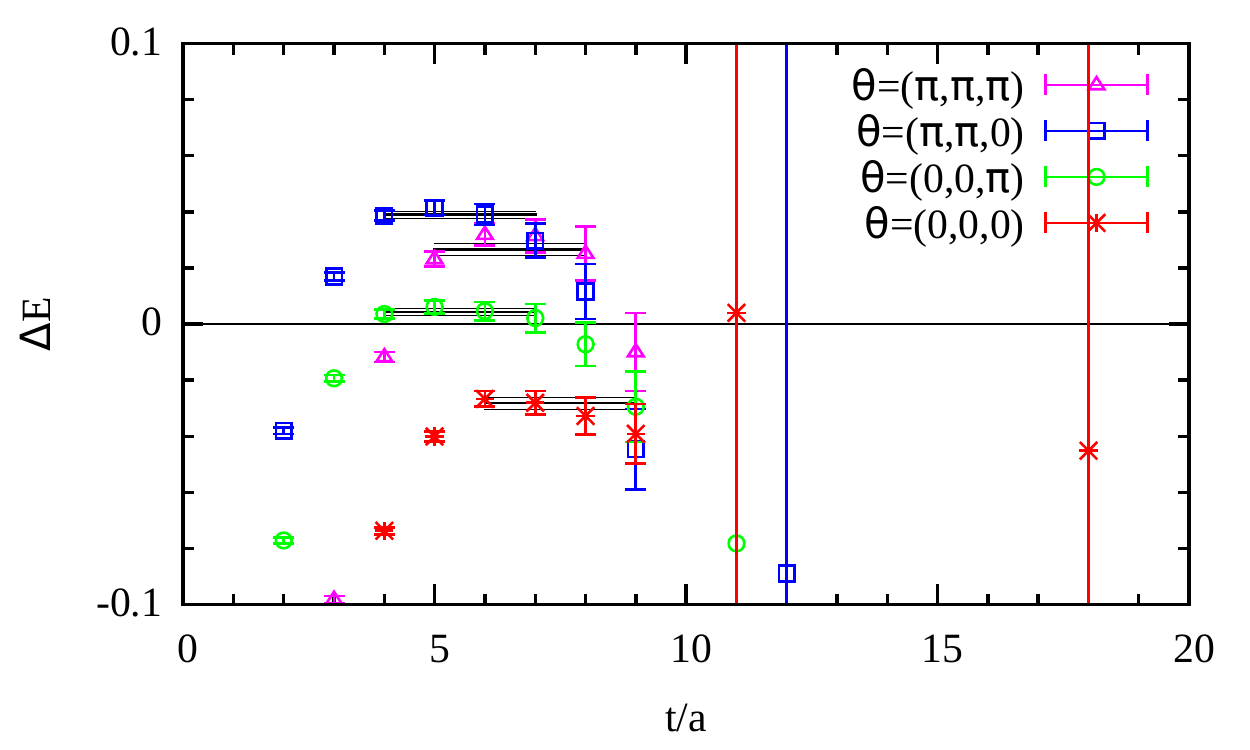}
  \includegraphics[scale=0.6]{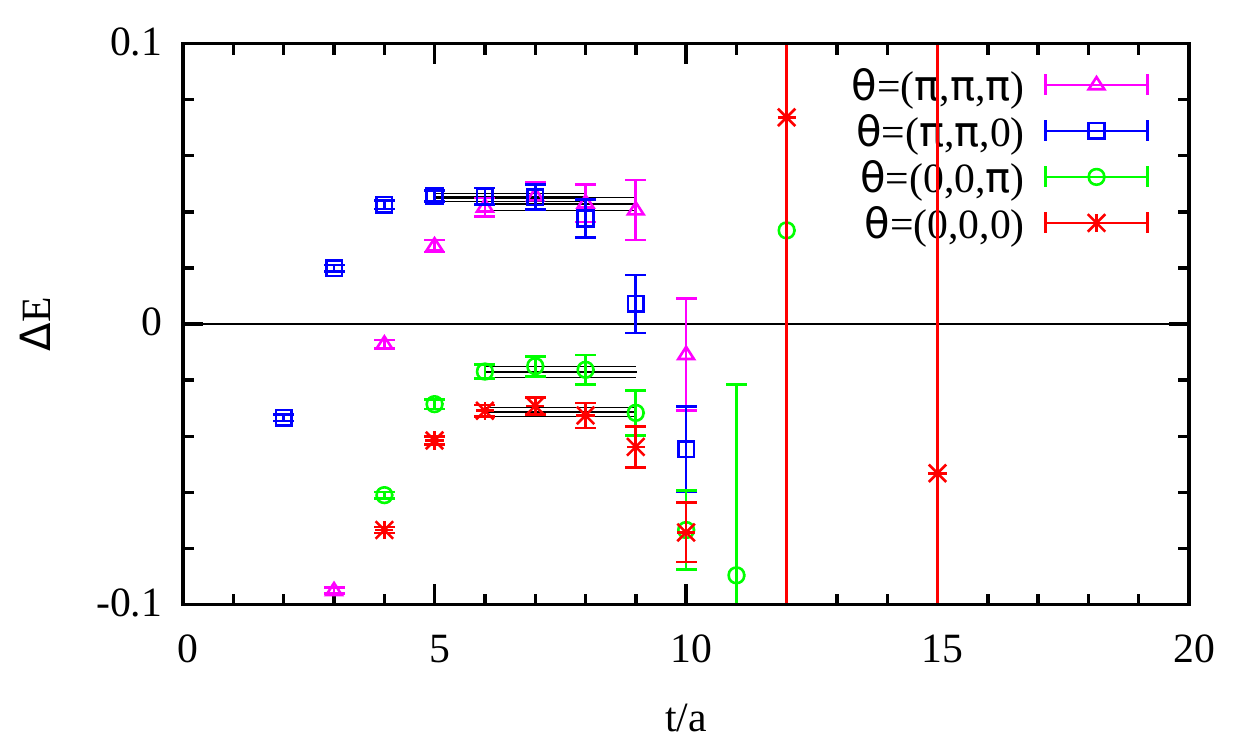}
  \includegraphics[scale=0.6]{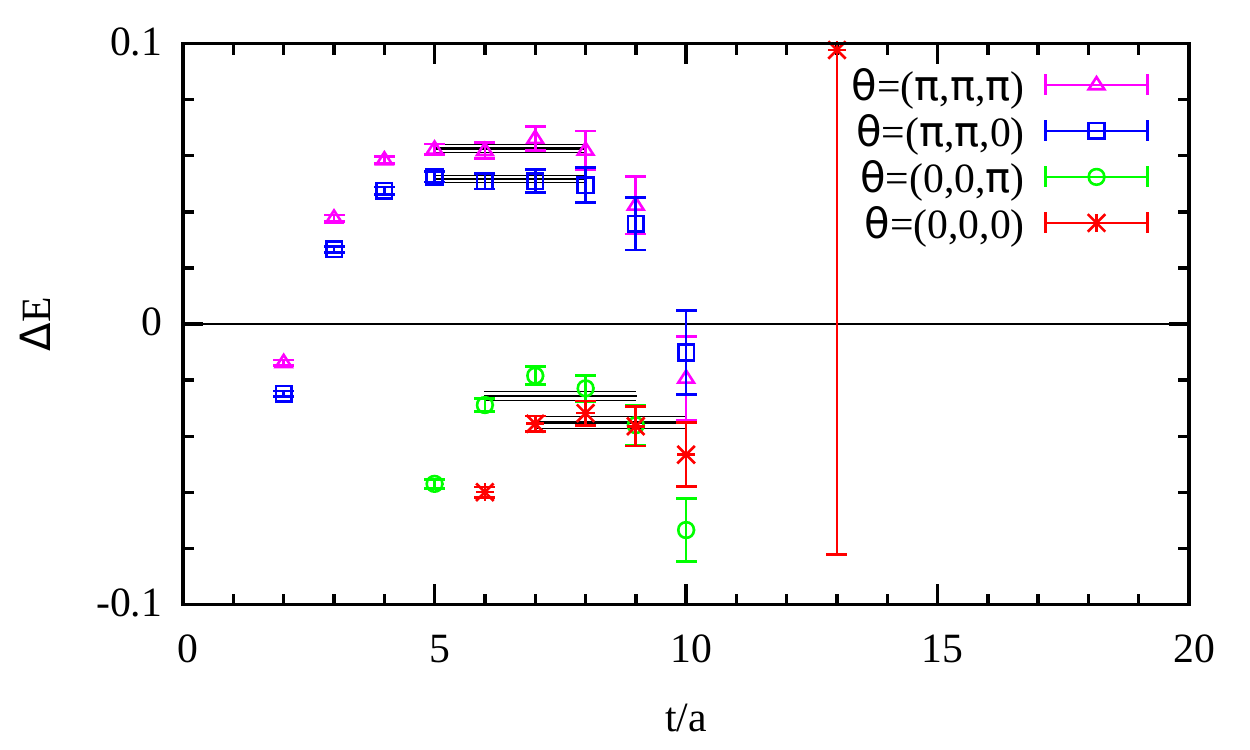}
  \caption{(color online) Summary of ground state eigenvalues $\Delta E_i$ at various twisting angles for three ensembles, from top to bottom for Ensemble I, II and III.}
  \label{Pic:Ground-energy-plateau-of-A1-channel}
\end{figure}

 Choices for the group reduction rules and momentum modes for different twist angles
 of scalar channel are listed in Table~\ref{Table:momentum-mode-number-in-A1-channel}.
\begin{table}[htb]
\begin{ruledtabular}
\centering
\caption{Information about group reduction rules and momentum modes for different twist angles in scalar channel.}
\begin{tabular}{ccccc}
\btheta    &$\mathbf{0}$    &$(0,0,\pi)$  &$(\pi,\pi,0)$  &$(\pi, \pi, \pi)$ \\
\hline
Symmetry   & $O_h$          & $D_{4h}$    & $D_{2h}$      & $D_{3d}$   \\
irreps     &  $A_1$         &  $A_1$      &  $A_1$        &  $A_1$  \\
Number of $\mathbf{k}_\alpha$     &3,2     & 3    &2     &3    \\
\end{tabular}
\label{Table:momentum-mode-number-in-A1-channel}
\end{ruledtabular}
\end{table}
 Initially three momentum modes are taken for all twisted angles. However, for the choice of $\btheta=(\pi,\pi,0)$,
 some choices of $t_0$ generates numerical instabilities which leads us to solve the smaller $2\times 2$
 sub-matrix. Whenever possible, we have also checked whether the lowest eigenvalues obtained from $3\times3$ and $2\times2$ sub-matrices are compatible with each other, and found that they are compatible within the limit of error.
 We conduct a search over a range of $t_0$  and the one that yields the minimum $\chi^2$ per degree of freedom is taken as the final result.
 As an illustration, the effective mass plots for energy shifts of the non-twisted case for three ensembles are shown in Fig.~\ref{Pic:Energy-plateau-of-A1-channel}, where one can see that plateaus can only be extracted from
 ground state. The ground state eigenvalues $\Delta E_i$ obtained from other twisting angles
 are similar and all of these are summarized in Fig.~\ref{Pic:Ground-energy-plateau-of-A1-channel}.
 while the numerical values are listed in Table~\ref{Table:deltaE-infomation-for-A1-channel}.


\begin{figure}[htb]
  \centering
  \includegraphics[scale=0.6]{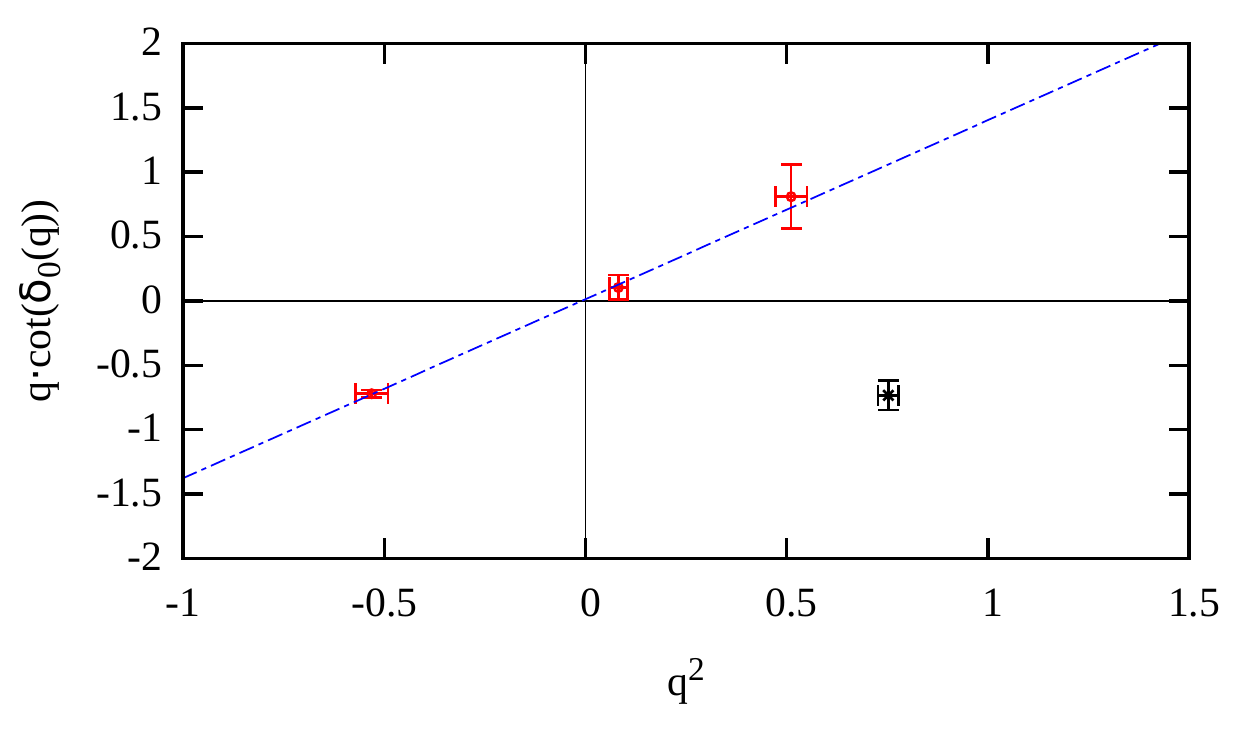}
  \includegraphics[scale=0.6]{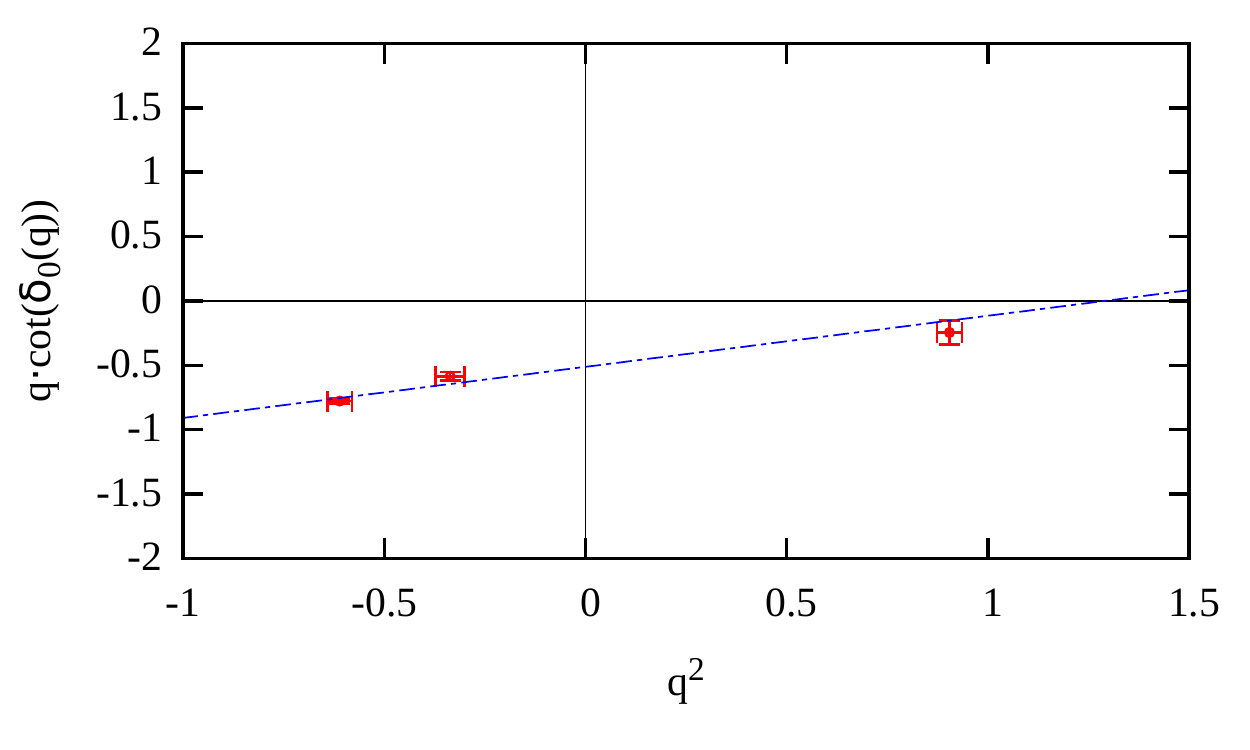}
  \includegraphics[scale=0.6]{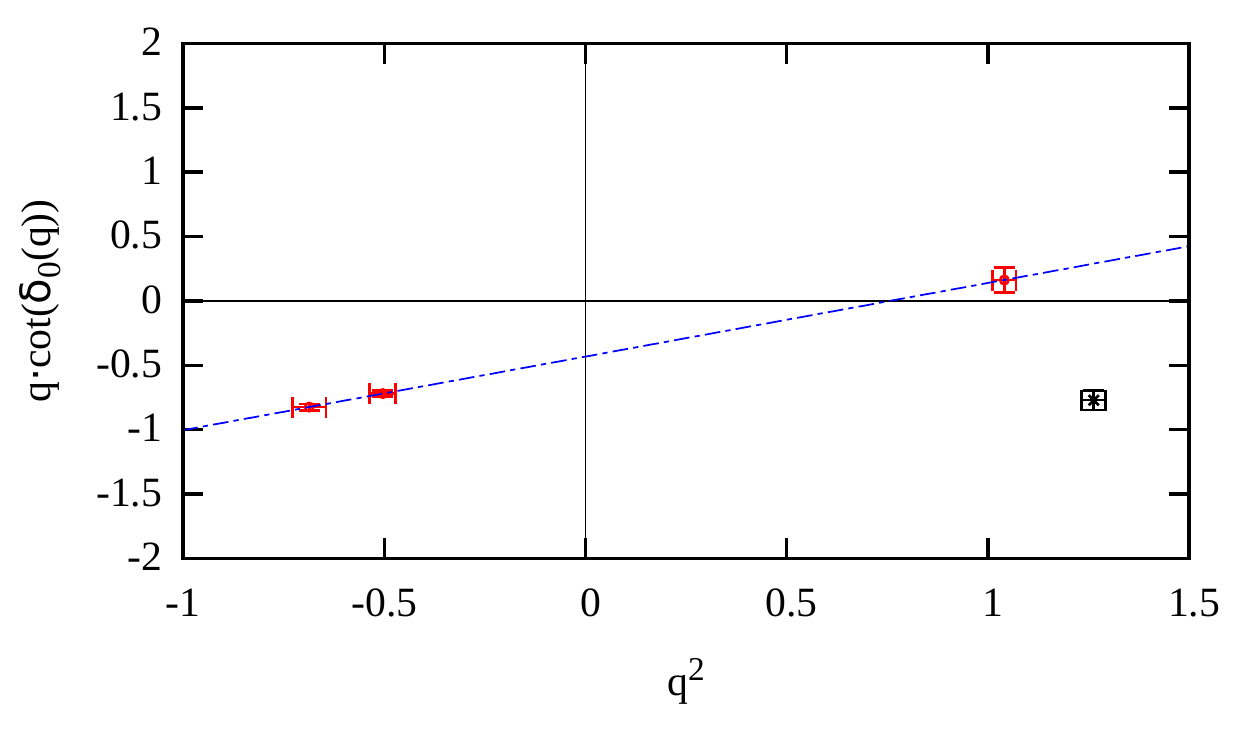}
  \caption{(color online) Fitting results based on the effective range expansion of Eq.~(\ref{Eqn:effective-range-expansion-q-s-wave}), from top to bottom for Ensemble I, II and III. Black star points are not taken into account in order to get a fitting with a relatively small $\chi^2$ and reasonable goodness-of-fit probability q. For Ensemble II, one point is out of range and thus invisible.}
  \label{Pic:Effective-range-expansion-for-A1}
\end{figure}

\begin{table*}[htb]
\begin{ruledtabular}
\caption{Simulation results for the scalar channel. The 3rd and 4th column gives the dimension of correlation matrix
and the reference time slice $t_0$. The fit range $[t_{\min},t_{\max}]$ from which we extract the values of
$\Delta E$ are also listed with $\chi^2/d.o.f$ in the next column. These ranges are relevant for the estimations
of $q^2$ and $m_{00}$, as both central values and errors are obtained from jackknifed samples and thus the errors are statistical.}
\begin{tabular}{ccccccccc}
\rm{Ensemble}         &$\btheta$   &Dim.   &Refts.  &Fit range &$\chi^2/d.o.f$  &$\Delta E$  &$q^2$   &$m_{00}$  \\
\hline
\multirow{4}{*}{\rm{I}} &$\mathbf{0}$ &  $3$   & 5 & [6, 9]  & 0.60 & -0.0282(21) &  -0.531(40) & -0.721(30)  \\
               &$(0, 0, \pi)$         &  $3$   & 6 & [4, 7]  & 0.35 & 0.0043(12) & 0.082(22)   &  0.105(95)  \\
               & $(\pi, \pi, 0)$      &  $2$   & 6 & [4, 7]  & 1.09 & 0.0390(13) &  0.753(26)     & -0.732(114) \\
               &$(\pi, \pi, \pi)$     &  $3$   & 5 & [5, 8]  & 1.37 & 0.0266(20) & 0.511(39)      &  0.811(248)  \\
\hline
\multirow{4}{*}{\rm{II}} &$\mathbf{0}$ &  $3$   & 5 & [6, 9]  & 1.15 & -0.0313(16)    &  -0.611(31) & -0.776(21)  \\
               &$(0, 0, \pi)$          &  $3$   & 5 & [6, 9]  & 1.22 & -0.0171(19) & -0.336(36)   &  -0.584(32)  \\
               & $(\pi, \pi, 0)$       &  $2$   & 6 & [5, 8]  & 0.43 & 0.0452(15) &  0.905(31) & -0.243(93) \\
               &$(\pi, \pi, \pi)$      &  $3$   & 5 & [6, 9]  & 0.17 & 0.0428(24) & 0.857(49)   &  -3.17(1.34)  \\
\hline
\multirow{4}{*}{\rm{III}} &$\mathbf{0}$ &  $2$   & 3 & [7, 10]  & 0.55 & -0.0351(22)    &  -0.687(42) & -0.825(26) \\
               &$(0, 0, \pi)$           &  $3$   & 4 & [6, 9]   & 3.26 & -0.0257(17) & -0.504(32)   &  -0.719(23)  \\
               & $(\pi, \pi, 0)$        &  $2$   & 6 & [5, 8]   & 0.13 & 0.0517(14) &  1.041(29) & 0.163(96) \\
               &$(\pi, \pi, \pi)$       &  $3$   & 6 & [5, 8]   & 0.26 & 0.0626(14) & 1.263(30)   &  -0.771(75)  \\
\end{tabular}
\label{Table:deltaE-infomation-for-A1-channel}
\end{ruledtabular}
\end{table*}

\subsubsection{$\bar{D}_1D^{*}$ scattering in s-wave channel}

 After the extraction of $\Delta E_i$, we can use the effective range expansion of Eq.~(\ref{Eqn:effective-range-expansion-q}) to extract the parameters.
 For s-wave scattering of $l=0$, the equation reads,
\begin{eqnarray}
 {q\cot\delta_0 (q^2)}=B_0+{1\over2} R_{0} q^2 +\cdots,
 \label{Eqn:effective-range-expansion-q-s-wave}
\end{eqnarray}
 where the l.h.s of this equation is calculated by Eq.~(\ref{Eqn:Luescher-Formula-m00}).
 The fitting results are illustrated in Fig.~\ref{Pic:Effective-range-expansion-for-A1}.
 Black star points in Fig.~\ref{Pic:Effective-range-expansion-for-A1} are left out
 in the final fitting procedure as their inclusion will hike up the final $\chi^2$
 of the fitting tremendously.

 The fitting results of $B_0$ and $R_0/2$ and the corresponding $\chi^2/\rm{dof}$
 for fitting results of different ensemble are listed in Table~\ref{Table:Scattering-parameters-for-A1}.
 Also listed in the last two rows are the physical values for the scattering parameters.

\begin{table}[htb]
\begin{ruledtabular}
\caption{Fitting results for the scattering length and effective range in the $A_1$ channel.}
\begin{tabular}{ccccc}
Ensemble      &     I        &   II         & III         \\
\hline
$B_0$         &  0.014(74)   & -0.511(33)   & -0.431(37)  \\
$R_0/2$       & 1.390(163)   & 0.397(60)    & 0.572(60)   \\
$\chi^2/\rm{dof}$  &  0.17        &   4.85       & 0.0016      \\
\hline
$a_0(\rm{fm})$     & 23.55(120.57)  & -0.66(4)    & -0.79(6)   \\
$r_0(\rm{fm})$     & 0.94(11)       & 0.27(4)     & 0.39(4)    \\
\end{tabular}
\label{Table:Scattering-parameters-for-A1}
\end{ruledtabular}
\end{table}

 As there are not good chiral behavior for the scattering parameters (except for $a_0$, but its value in Ensemble I is nearly divergent and no reasonable extrapolation can be conducted here), we would like only to keep the individual values for each case. But we can see that the values of $r_0$ for three ensembles are all much smaller than the size of the lattice, indicating that using of effective range expansion here is reasonable.

\subsubsection{Possibility of shallow bound state in $A_1$ channel}

 \begin{figure}[htb]
  \centering
  \includegraphics[scale=0.6]{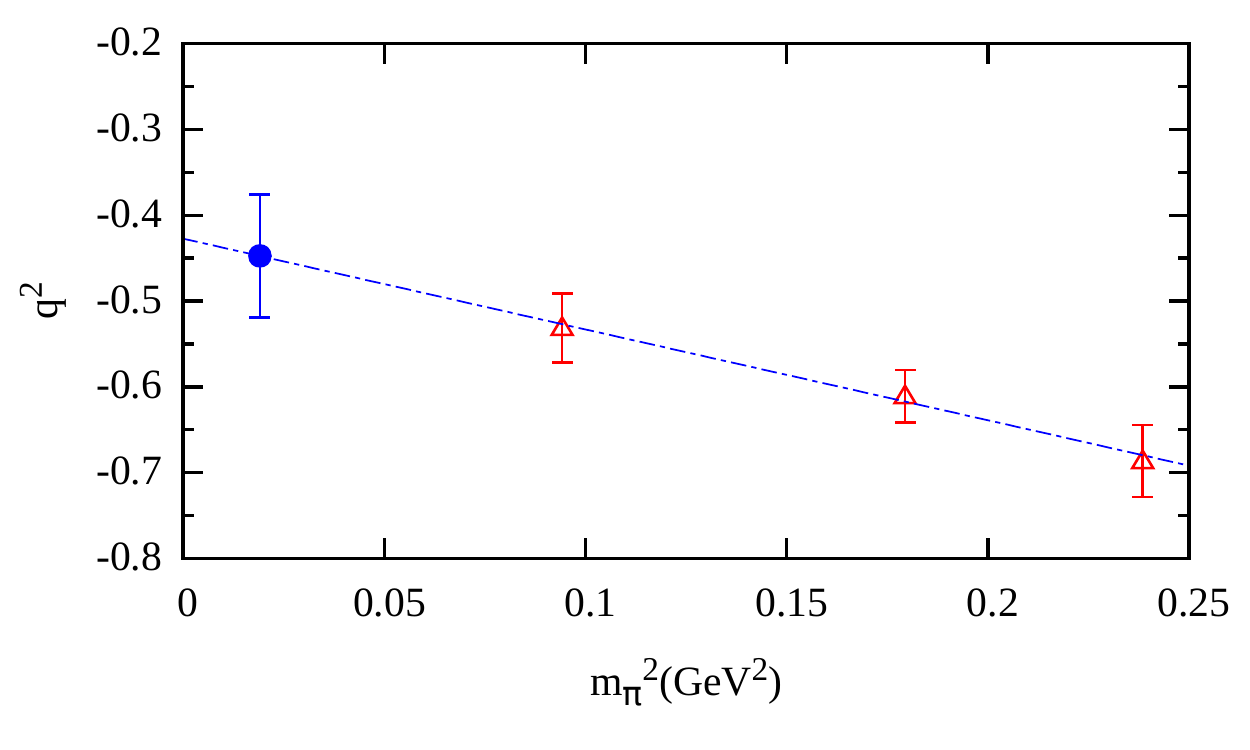}
  \includegraphics[scale=0.6]{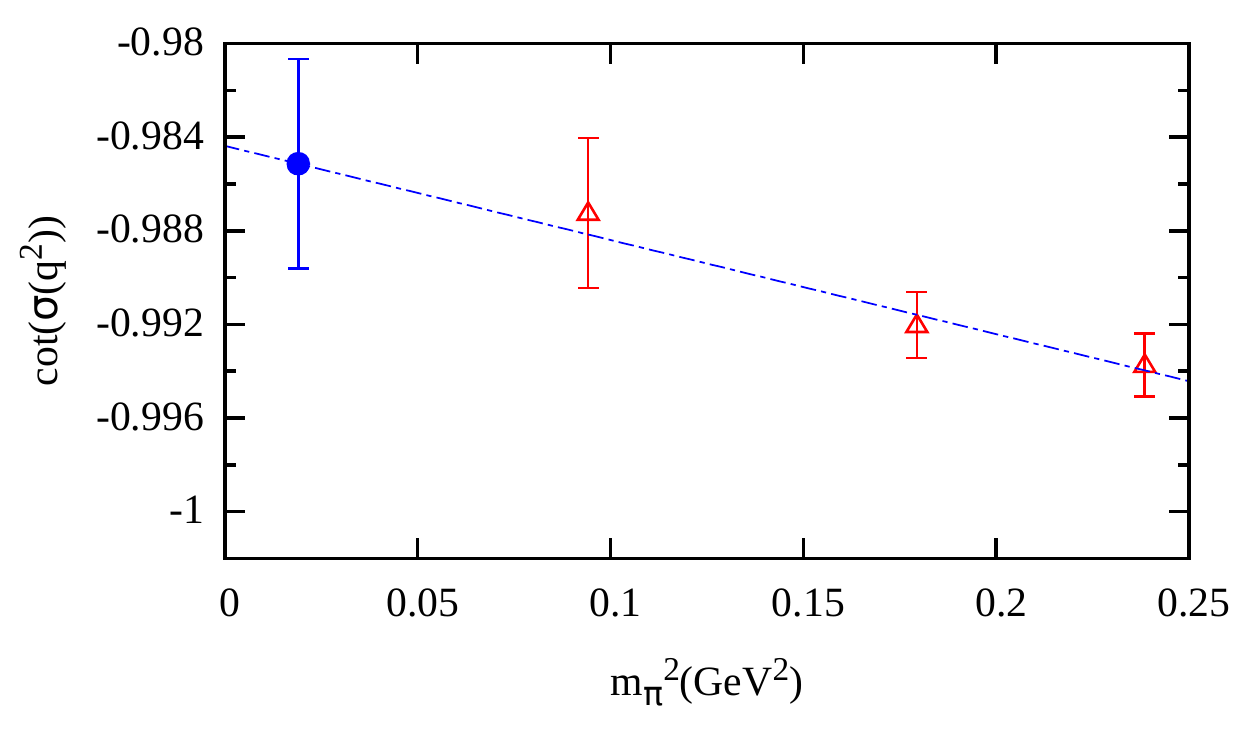}
  \caption{(color online) Chiral limit of dimensionless momentum $q^2$ (upper panel) and $\cot\sigma(q)$ (lower panel) of ground state for three ensembles in $A_1$ channel, with fitting $\chi^2/\rm{dof}=0.08, ~0.21$ alternatively.}
  \label{Pic:q2-chiral-extrapolation-A1}
\end{figure}
 To explore the possibility of a bound state in $A_1$ channel, we will use the formalism given in Sec.~(\ref{Subsec:bound-state-formation}) which tells us, in order to have a genuine bound state,
 the value of $q^2$ should be negative and $q^2\rightarrow -\infty$ as $L\rightarrow\infty$.
 The values for the lowest $q^2$ in $A_1$ channel come out to be in the range $[-0.7,-0.5]$
 which are indeed negative.
 Compared with earlier quenched results for the lowest $q^2$ of different volume, ranging between
 $[-0.07,-0.02]$ (see Table~III in Ref.~\cite{Meng:2009qt}),
 we can see that the absolute values of $q^2$ are roughly increased by an order of magnitude,
 which means the interaction between the two mesons indeed becomes stronger.
 We can also proceed to evaluate the corresponding value of $\cot\sigma(q)$ from Eq.~(\ref{Eqn:Luescher-Formula-m00-minusq2}).
 These results, which are rather close to the value of $-1$,
 are also tabulated in Table~\ref{Table:cot-sigmaq-for-A1}.
 It is interesting to inspect the chiral behavior of the lowest $q^2$ and
 the values of $\cot\sigma(q)$ from our three ensembles.
 Admittedly we have only three different pion mass values which are also quite far
 away from the chiral limit, a naive extrapolation linear in $m^2_\pi$ is
  still performed for the lowest $q^2$ and $\cot\sigma(q)$ respectively.
 These are illustrated in Fig.~\ref{Pic:q2-chiral-extrapolation-A1}.

\begin{table}
\begin{ruledtabular}
\caption{Results for the lowest $q^2$ and the corresponding values for $\cot\sigma(q)$ as given by Eq.~(\ref{Eqn:Luescher-Formula-m00-minusq2}) in the $A_1$ channel for three ensembles.
Corresponding statistical errors for the quantities are given in the parenthesis.
The last column gives the chiral extrapolation of $q^2$ and $\cot\sigma(q^2)$.}
\centering
\begin{tabular}{ccccc}
    $\rm{Ensemble}$   & $\rm{I}$    &  $\rm{II}$    &$\rm{III}$     & Chiral Limit   \\
    \hline
    $m_{\pi}$[GeV]    &  0.3070     &   0.4236      &  0.4884       &  0.1380        \\
    $q^2$             & -0.531(40)  &  -0.611(31)   & -0.687(42)    &  -0.447(71)    \\
    $\cot\sigma(q^2)$ & -0.9872(32) &  -0.9920(14)  & -0.9937(13)   &  -0.9851(45)              \\


\end{tabular}
\label{Table:cot-sigmaq-for-A1}
\end{ruledtabular}
\end{table}
 Even though the values of $\cot\sigma(q^2)$ are very close to $-1$ for the three ensembles, which seem to
 indicate the formation of a bound state in this channel, we have to point out that the chiral
 behavior of $q^2$ is moving upwards which is deviating from $\cot\sigma(q^2)=-1$ as $m^2_\pi$ is decreased.
 Bearing in mind that the value of $m_\pi L$ for the lightest point is somewhat small,
 one may worry that finite volume systematic effects are contaminating the data point at lower $m_\pi$ values.
 Therefore, we can only say that, at this stage our data cannot rule out the existence of a bound state
 in $A_1$ channel and a more careful study with different volumes is necessary to further clarify
 the situation.  So the bottom line is, with dynamical quarks into the simulation as opposed to
 the previous quenched study, the attraction between the two charmed mesons appears to be stronger.

\begin{table}[htb]
\centering
\caption{Information about group reduction rules and momentum numbers for different twist angles in vector channel.}
\begin{tabular}{ccccc}
\hline
\hline
\btheta    &$\quad\mathbf{0}\quad$    &$\quad(0,0,\pi)\quad$  &$\quad(\pi,\pi,0)\quad$  \\
\hline
Symmetry   & $O_h$          & $D_{4h}$    & $D_{2h}$      \\
irreps     &  $T_1$         &  $E$      &  $A_1$        \\
Number of $\mathbf{k}_\alpha$     &2     & 2    &2      \\
\hline
\hline
\end{tabular}
\label{Table:momentum-mode-number-in-T1-channel}
\end{table}
\subsection{Results for the Vector channel}
\label{Subsec:vector-channel-result}

 Similar to the scalar channel, we have checked the charge parity for
 $\epsilon = \pm 1$ as shown in Eq.~(\ref{Operator:two-particle-contents-T1}), and only
 found signal for negative charge parity with $\epsilon = 1$. Thus,
 following results are all for two-particle
 states with quantum number of $I^G(J^{PC}) = 1^+(1^{+-})$.

\subsubsection{Two-particle energy spectra}

The momentum number and twist angle with corresponding group reduction used for the vector channel are listed in the Table~\ref{Table:momentum-mode-number-in-T1-channel}.

\begin{figure}[htb]
  \centering
  \includegraphics[scale=0.6]{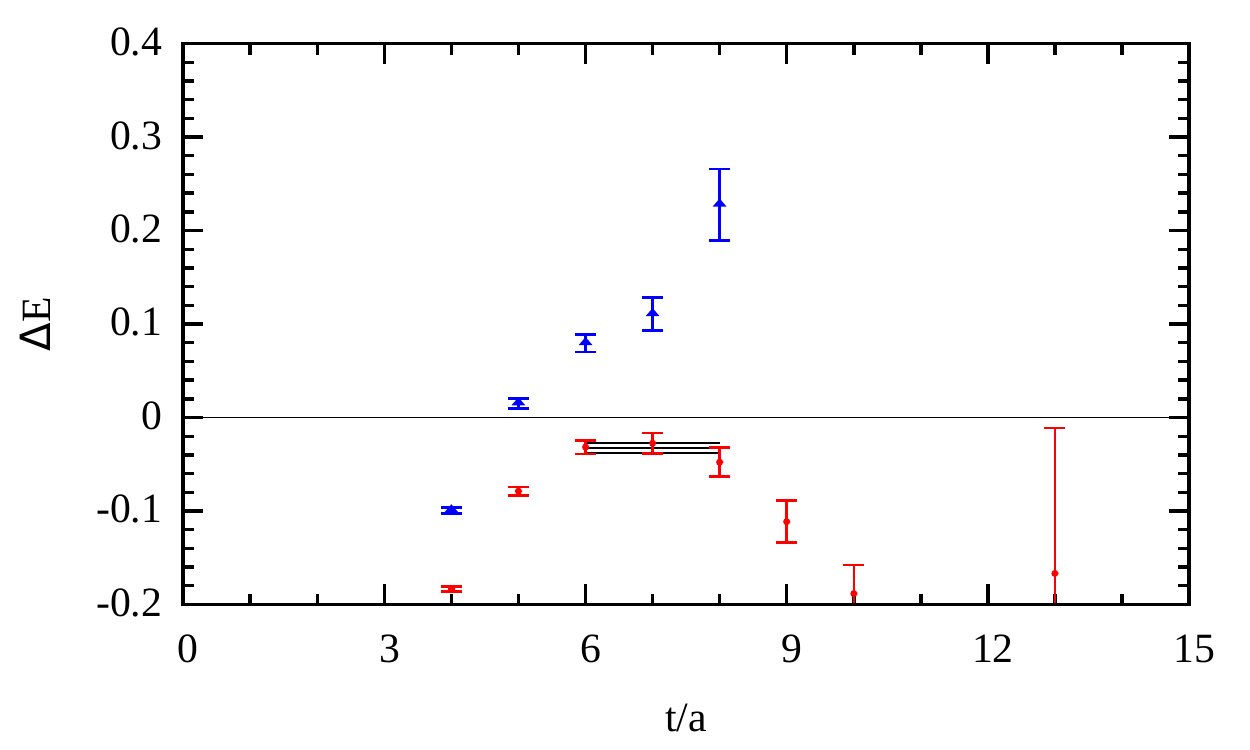}
  \includegraphics[scale=0.6]{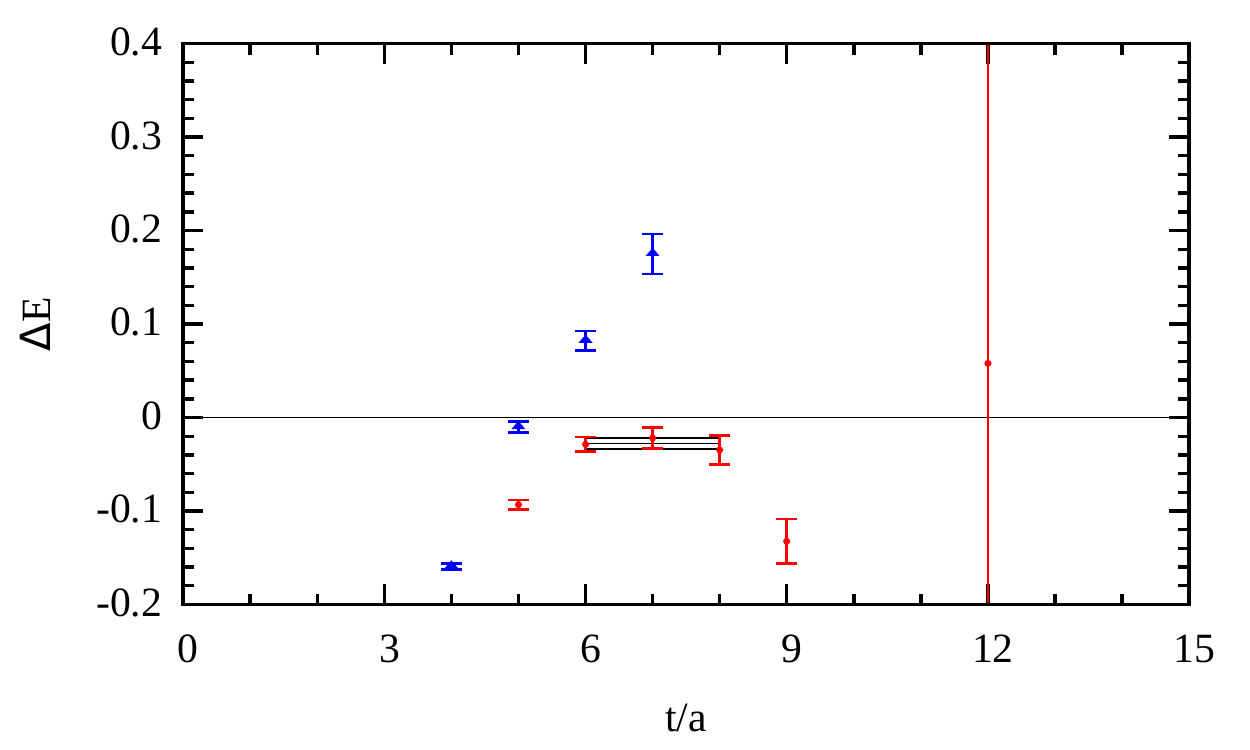}
  \includegraphics[scale=0.6]{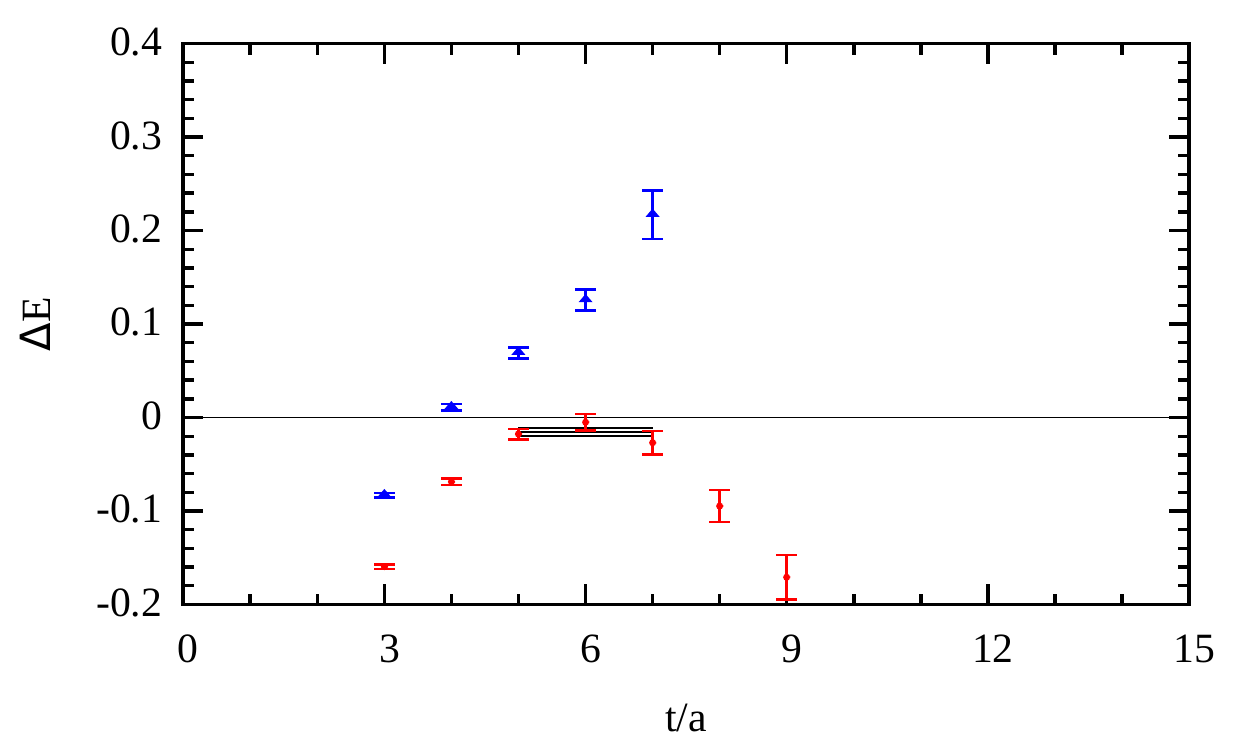}
  \caption{(color online) Effective mass plots for the energy shift $\Delta E_\alpha$ for Ensemble II for different twisting angles $\btheta=\mathbf{0}$ (top), $\btheta=(0,0,\pi)$ (middle) and $\btheta=(\pi,\pi,0)$ (bottom). The two energy level stands for $\alpha = 1,2$, and the grey horizontal bars indicate the fitted values for $\Delta E_\alpha$'s within the fitting ranges.}
  \label{Pic:Energy-plateau-of-T1-channel}
\end{figure}
\begin{figure}[htb]
  \centering
  \includegraphics[scale=0.6]{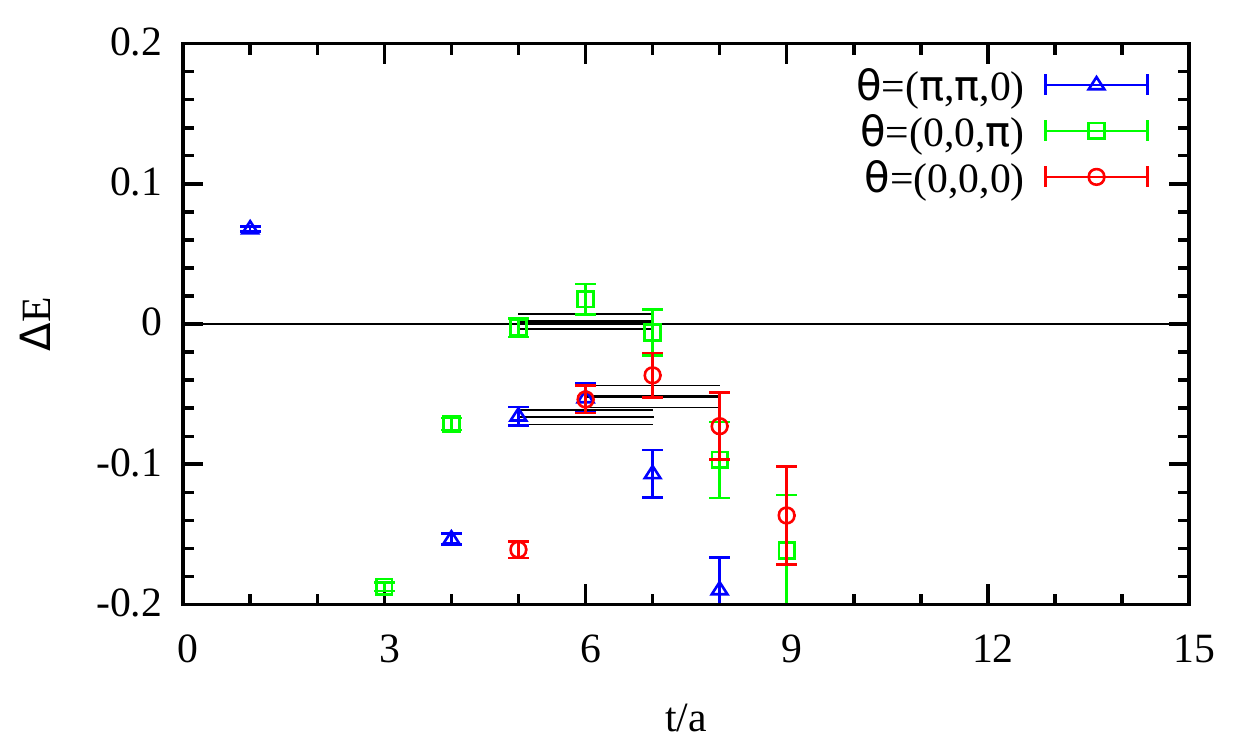}
  \includegraphics[scale=0.6]{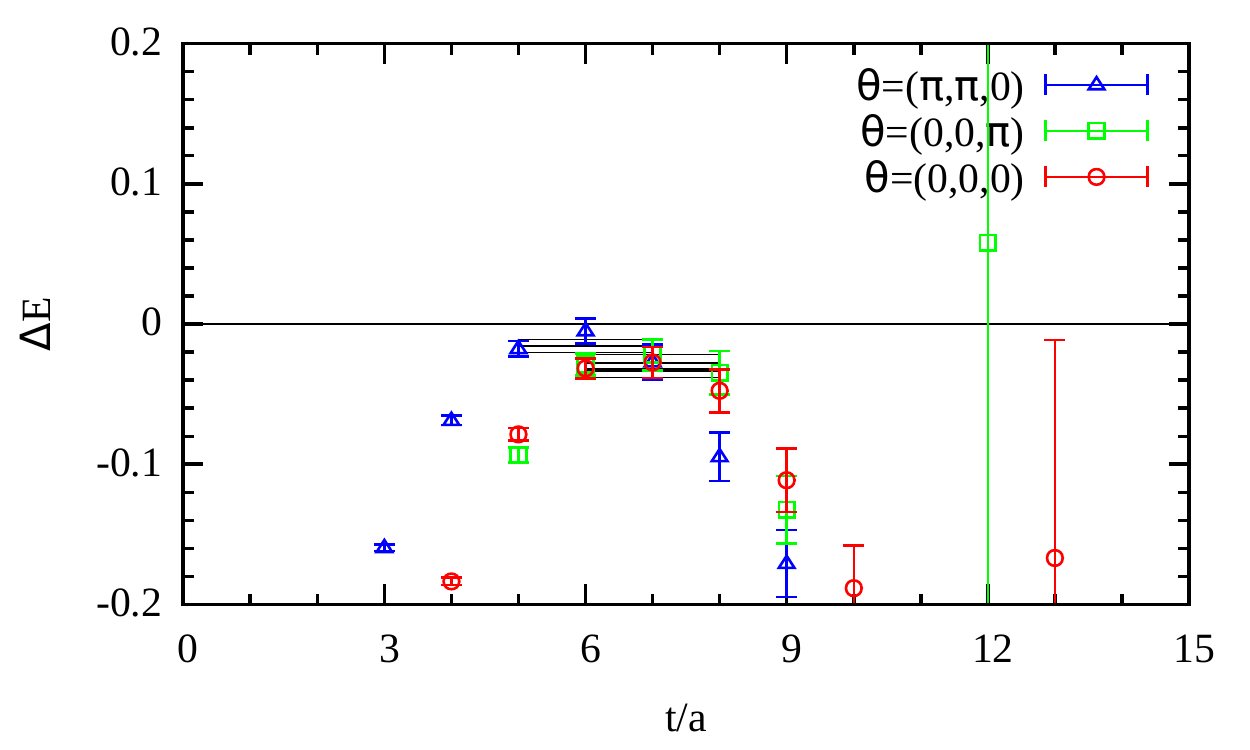}
  \includegraphics[scale=0.6]{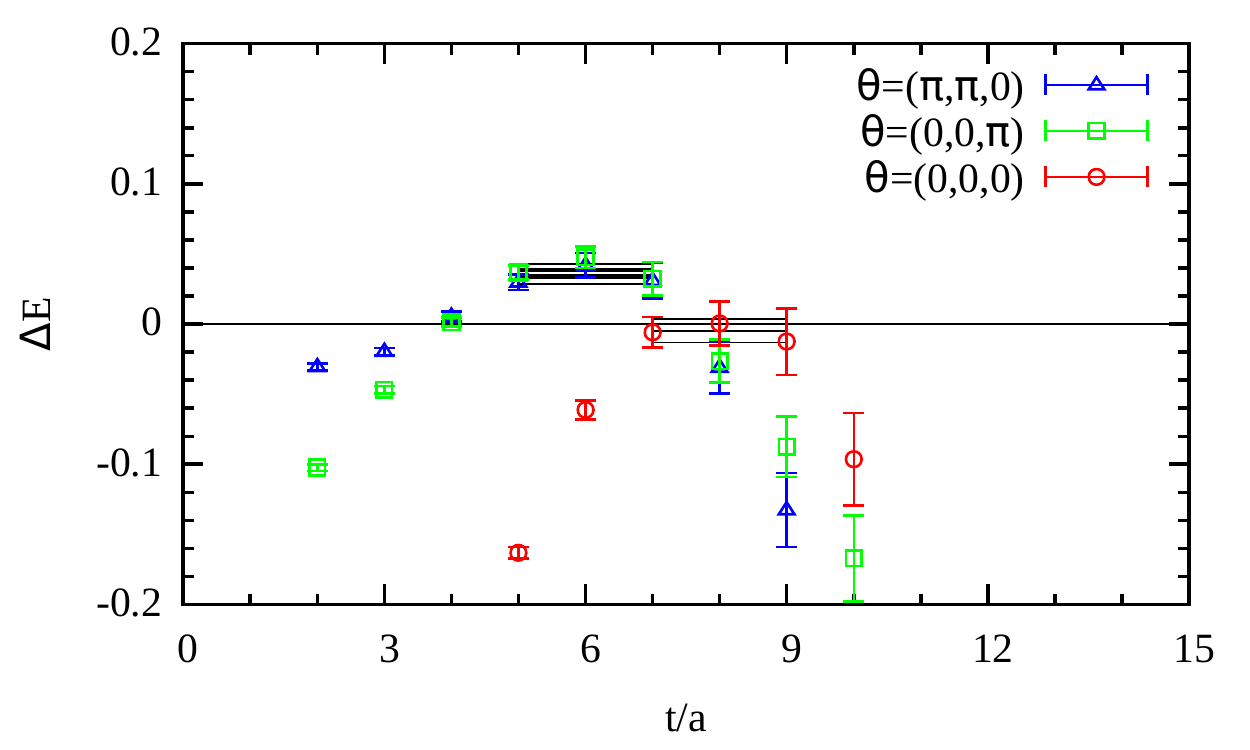}
  \caption{(color online) Ground eigenvalues got from every correlation matrix for three ensembles, from top to bottom for Ensemble I, II and III.}
  \label{Pic:Ground-energy-plateau-of-T1-channel}
\end{figure}

As there are no zero-momentum mode in the construction of operators for vector channel as discussed in Sec.~\ref{Subsec:operator-two-particle-contents-T1}, we only take two momentum for $T_1$ channel, which corresponds $\alpha=1,2$ for $\mathbf{k}_\alpha$.
As an illustration, the effective mass plots of Ensemble II for all three twisting angles are shown in Fig.~\ref{Pic:Energy-plateau-of-T1-channel}. It is seen that the signal is worse than that in the $A_1$ channel.
Normally we can only get a plateau that extends for about 3
 consecutive points and the errors are also quite large.
The energy shifts $\Delta E_i$'s for this case and for other twisting angles are summarized in Fig.~\ref{Pic:Ground-energy-plateau-of-T1-channel}. The numerical results for $\Delta E_i$'s
are collected in Table~\ref{Table:deltaE-infomation-for-T1-channel}.

\begin{table*}[htb]
\begin{ruledtabular}
\caption{Simulation results for the vector channel. Jackknifed samples are used to estimate center value of $\Delta E$, $q^2$ and $m_{11}$ alternatively and all the errors are statistical. Only the ground state corresponding to $\mathbf{k} = (0,0,1)$ is used, higher modes are neglected as no stable plateau can be extracted.}
\begin{tabular}{ccccccccc}
\rm{Ensemble}         &$\btheta$   &Irrep      &Refts &Fit range &$\chi^2/d.o.f$ &$\Delta E$ &$q^2$   &$m_{11}$  \\
\hline
\multirow{3}{*}{\rm{I}} &$\mathbf{0}$ &  $T_1$ &  2  & [6, 8]    & 0.87 & -0.0516(78) & -0.997(0.148) &  0.951(0.218) \\
               &$(0, 0, \pi)$        &  $E$    &  5  & [5, 7]    & 1.40 &  0.0019(54) &  0.036(0.105) & -0.080(0.042) \\
              & $(\pi, \pi, 0)$     &  $A_1$   &  4  & [5, 7]    & 3.68 & -0.0664(53) & -1.277(0.099) &  1.389(0.165) \\
\hline
\multirow{3}{*}{\rm{II}} &$\mathbf{0}$ &  $T_1$&  4  & [6, 8]    & 0.60 & -0.0326(55) & -0.656(0.110) &  0.511(0.133) \\
               &$(0, 0, \pi)$        &  $E$    &  3  & [6, 8]    & 0.25 & -0.0277(59) & -0.557(0.118) &  0.398(0.133) \\
              & $(\pi, \pi, 0)$    &  $A_1$    &  5  & [5, 7]    & 1.20 & -0.0157(43) & -0.318(0.087) &  0.130(0.055) \\
\hline
\multirow{3}{*}{\rm{III}} &$\mathbf{0}$ &  $T_1$& 2  & [7, 9]    & 0.11 & -0.0049(83) & -0.099(0.168) & -0.012(0.095) \\
               &$(0, 0, \pi)$        &  $E$     & 6  & [5, 7]    & 0.92 &  0.0391(39) &  1.514(0.059) & -2.212(0.675) \\
              & $(\pi, \pi, 0)$    &  $A_1$     & 6  & [5, 7]    & 0.69 &  0.0331(44) &  1.592(0.069) & -2.315(1.772) \\
\end{tabular}
\label{Table:deltaE-infomation-for-T1-channel}
\end{ruledtabular}
\end{table*}

\subsubsection{$\bar{D}_1D^{*}$ scattering in p-wave channel}

\begin{figure}[htb]
  \centering
  \includegraphics[scale=0.6]{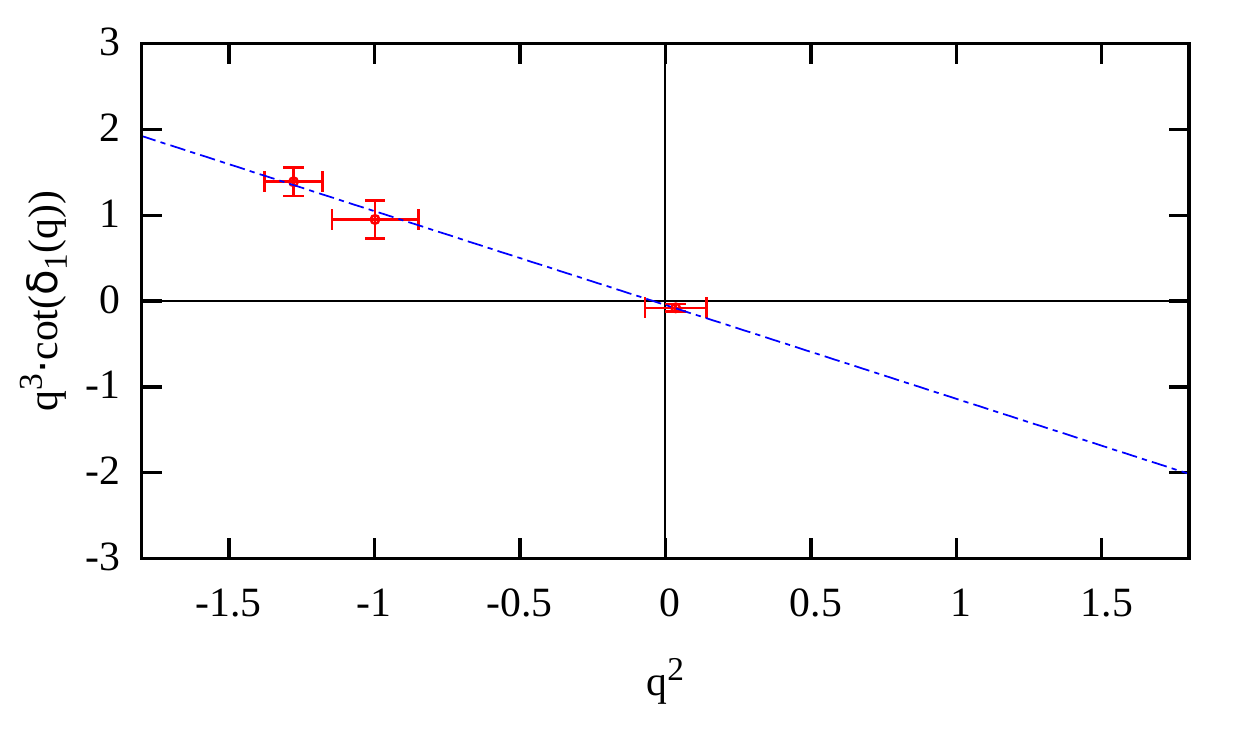}
  \includegraphics[scale=0.6]{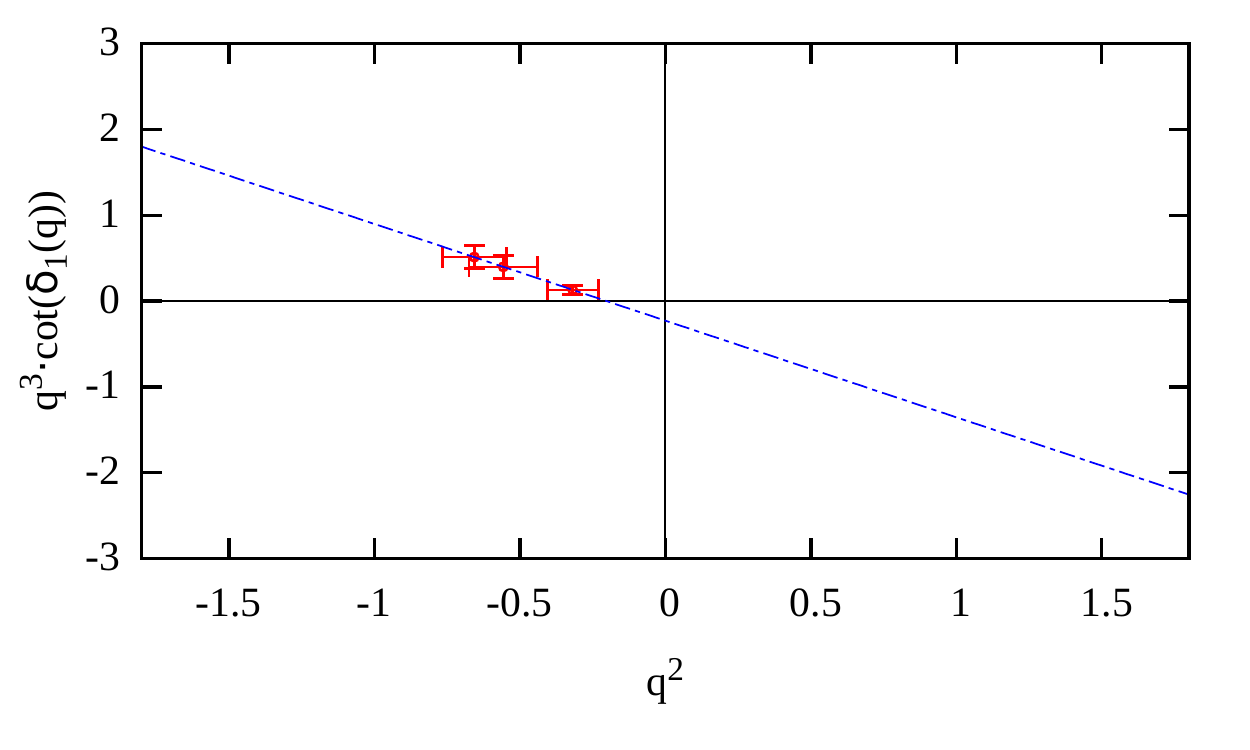}
  \includegraphics[scale=0.6]{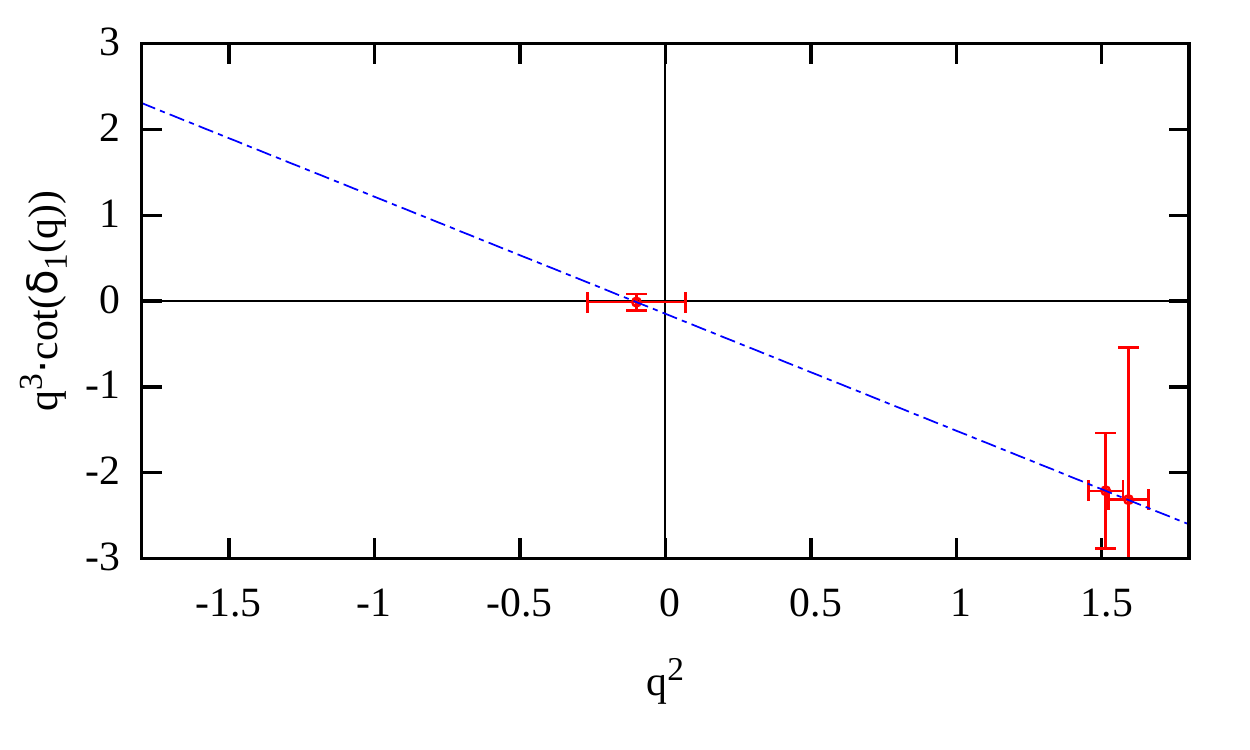}
  \caption{(color online) Fitting results based on the effective range expansion
  of Eq.~(\ref{Eqn:effective-range-expansion-q-p-wave}), from top to bottom for Ensemble I, II and III.}
  \label{Pic:Effective-range-expansion-for-T1}
\end{figure}

 Similar to the procedure in s-wave channel, the effective range expansion
 Eq~.(\ref{Eqn:effective-range-expansion-q}) is utilized to extract the parameters.
 For $p$-wave scattering of $l=1$, the equation can be expressed as
\begin{eqnarray}
 {q^3\cot\delta_1 (q^2)}=B_1+{1\over2} R_{1} q^2 +\cdots,
 \label{Eqn:effective-range-expansion-q-p-wave}
\end{eqnarray}
 where the l.h.s of this equation is calculated by equations given
 in Table~\ref{Table:m-function-for-SP}. The results are illustrated as in Fig.~\ref{Pic:Effective-range-expansion-for-T1}. Then a fit is performed
 and the fitting results for $B_1$ and $R_1/2$ and their corresponding $\chi^2/\rm{dof}$
 are listed in Table~\ref{Table:Scattering-parameters-for-T1}, while the last two rows are the recovered
 physical values of scattering parameters by the definition in Eq.~(\ref{Eqn:effective-range-expansion-q}).

\begin{table}[htb]
\begin{ruledtabular}
\caption{Fitting results for the scattering length and effective range in the $T_1$ channel.}
\begin{tabular}{ccccc}
Ensemble       &      I          &       II         &   III          \\
\hline
$B_1$          & -0.045(121)   & -0.228(328)    &  -0.147(241) \\
$R_1/2$        & -1.093(165)   & -1.127(588)    &  -1.364(427) \\
$\chi^2/\rm{dof}$ &   0.16       &  0.000034        &   0.000004     \\
\hline
$a_1(\rm{fm}^3)$     &  -0.865(2.305)   & -0.172(0.247)   & -0.266(0.435) \\
$r_1(\rm{fm}^{-1})$     &  -6.441(0.973)   & -6.636(3.467)   &  -8.032(2.516) \\
\end{tabular}
\label{Table:Scattering-parameters-for-T1}
\end{ruledtabular}
\end{table}

 Similar to the situation in $A_1$ channel, the scattering volume $a_1$ also suffers
 from the huge error for Ensemble. I and no reasonable chiral extrapolation can be conducted. However,
 the chiral behavior of $r_1$ seems to be good.

\subsubsection{Possibility of shallow bound state in $T_1$ channel}

 To explore the possibility of a bound state in $T_1$ channel, we follow the same procedure as in $A_1$ channel.
 We again use the formalism given in Sec.~(\ref{Subsec:bound-state-formation}) for negative $q^2$ of the lowest energy level. We can also compute the value of $\cot\sigma(q^2)$ at the lowest $q^2$ for each ensemble.
 They turn out to be close to $-1$, signaling a possible bound state.
 \begin{figure}[htb]
  \centering
  \includegraphics[scale=0.6]{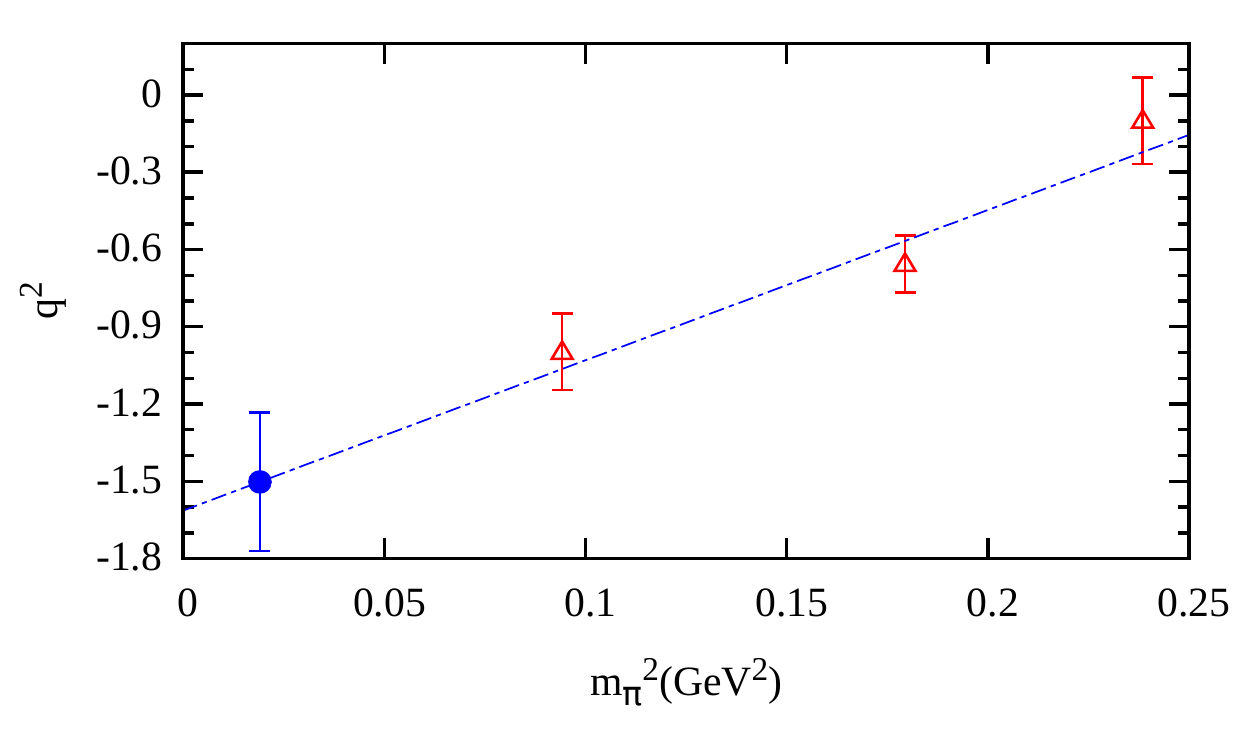}
  \includegraphics[scale=0.6]{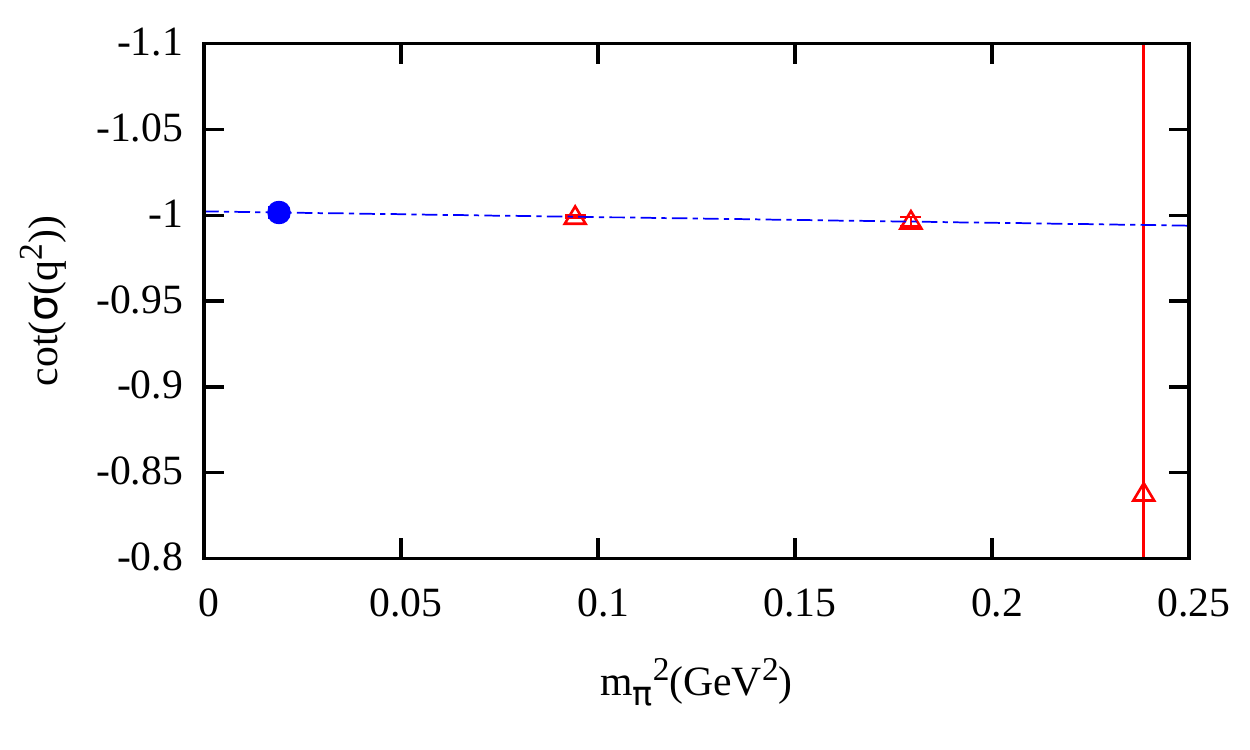}
  \caption{(color online) Chiral limit of dimensionless momentum $q^2$(upper panel) and $\cot\sigma(q^2)$(lower panel) of ground state for three ensembles in $T_1$ channel, with fitting $\chi^2/\rm{dof}=1.39,~0.07$ alternatively.}
  \label{Pic:q2-chiral-extrapolation-T1}
\end{figure}
 We can even inspect the chiral behavior of the lowest $q^2$ and $\cot\sigma(q^2)$
 which is shown in Fig.~\ref{Pic:q2-chiral-extrapolation-T1}.
 The results for the lowest (negative) $q^2$ and the corresponding values of $\cot\sigma(q)$ as computed from Eq.~(\ref{Eqn:Luescher-Formula-m00-minusq2}) are listed in Table~\ref{Table:cot-sigmaq-for-T1}.

\begin{table}
\begin{ruledtabular}
\caption{Results for the lowest $q^2$ and the corresponding values for $\cot\sigma(q)$ as given by Eq.~(\ref{Eqn:Luescher-Formula-m00-minusq2}) in $T_1$ channel for three ensembles.
 Corresponding statistical errors for the quantities are given in the parenthesis.
 The last column gives the chiral extrapolation of $q^2$ and $\cot\sigma(q^2)$.}
\centering
\begin{tabular}{ccccc}
    $\rm{Ensemble}$   & $\rm{I}$    &  $\rm{II}$    &$\rm{III}$     & Chiral Limit    \\
    \hline
    $m_{\pi}$[GeV]    &  0.3070     &   0.4236      &  0.4884       &  0.1380         \\
    $q^2$             & -0.997(148) &  -0.656(110)  & -0.099(167)   &  -1.502(269)    \\
    $\cot\sigma(q^2)$ & -0.9991(6)  &  -0.9963(26)  & -0.8379(5976) &  -1.0016(33)     \\


\end{tabular}
\label{Table:cot-sigmaq-for-T1}
\end{ruledtabular}
\end{table}

 We see that the chiral behavior of lowest $q^2$ in $T_1$ channel is opposite to that in $A_1$ channel,
 leading to a much deeper $q^2$ value in the chiral limit, which might mean that there is a bound
 state forming in this channel. However, we only have one volume for the three ensembles,
 and therefore are unable to perform the finite volume extrapolation within this formalism.
 Further exploration with different volumes should be conducted in order to reach a more definite conclusion.

\section{Conclusions}
\label{Sec:conclusion}

 In this paper, we have performed an exploratory lattice study for the
 low-energy scattering of the $(\bar{D}_1D^{*})^+$ two-particle system in both s-wave($A_1$)
 and p-wave($T_1$) channel, corresponding to the quantum numbers of
 $I^GJ^{PC} = 1^+0^{--}$ and $I^GJ^{PC} = 1^+1^{+-}$ respectively.
 Assuming that close to the threshold the system is dominated by elastic scattering,
 we used the standard L\"uscher formalism to study their interactions.
 It is found that in both channels, the interaction between the two charmed mesons
 is attractive in nature. There are also indications that they might form bound states but
 a definite conclusion can only be made when more systematic studies with different volumes are performed.
 Positive charge parity channels are also investigated with no signals found.

 The calculation is based on the $N_f = 2$ twisted mass fermion configurations of size $32^3\times64$ with a lattice spacing of about $0.067\rm{fm}$. Three ensembles of different pion mass with $m_{\pi}L = 3.31, 4.57, 5.27$ are
 utilized to investigate the pion mass dependence of various physical quantities in the simulation.
 In order to enhance the momentum resolution around the two-particle threshold,
 twisted boundary conditions are utilized together with the conventional periodic boundary conditions.
 We only take the twist angle that are integral multiples of $\pi$ and thus avoid the mixing of partial waves with opposite parity.  These techniques lead to a perfect dispersion relation
 for the vector meson; for the axial vector meson, albeit the much noisier
 correlation function, reasonable results are obtained.

 For the two-particle scattering in $A_1$ channel,
 the results in this paper update our former quenched results.
 The attraction between the two charmed mesons appears to be stronger compared with
 the quenched case which is represented by a much more negative value of the lowest $q^2$.
 We have also checked the possibility of bound state formation in $A_1$ channel by checking
 the quantity $\cot\sigma(q^2)$ within L\"uscher's formalism.
 For all three ensembles, the values of $\cot\sigma(q^2)$ turn out to be rather close to $-1$, which is
 the value signaling a bound state. However, due to possible finite volume contaminations, we
 still cannot draw a definite conclusion whether there is a bound state in this channel but our results
 cannot rule it out either.

 In the $T_1$ channel, similar conclusions are reached. By inspecting the lowest values
 of $q^2$ and the quantity $\cot\sigma(q^2)$, it is seen that the two mesons have attractive
 interaction and the value of $\cot\sigma$ is also compatible with a bound state.
 However, due to the relative poor signal to noise ratio and the possible finite volume
 contamination of the lightest pion mass point, it is still premature to draw any definite conclusions.

 Based on the discussion above, it is seen that, quite contrary to the charmed meson interaction below
 $4.2 \rm{GeV}$ where the interaction appears to be mostly repulsive in nature~\cite{Chen:2014afa, Chen:2015jwa, Prelovsek:2013xba, Prelovsek:2014swa, Lang:2015sba},
 interactions between a $(\bar{D}_1D^{*})^{\pm}$ two-particle system is attractive.
 The interaction is also stronger compared with the quenched case.
 In both channels (pseudo-scalar and axial vector),
 our lattice data show indications of a possible bound state below
 the threshold, though a much more careful multi-volume study should be performed before
 any definite conclusions can be made.
 We also hope this will shed some light to the nature of newly identified $Z(4430)$ structure.

\section*{Acknowledgments}
The authors would like to thank F.~K.~Guo, L.~M.~Liu, Y.~B.~Yang, U.~Meissner, A.~Rusetsky, C.~Urbach and B.~Knippschild for helpful discussions. The authors would also like to thank the European Twisted Mass Collaboration (ETMC) to allow us to use their gauge field configurations. Our thanks also go to Supercomputing Center of Chinese Academy of Science (SCCAS) and the Bejing Computing Center (BCC) where part of the numerical computations are performed. This work is supported in part by the National Science Foundation of China (NSFC) under the project No.11335001, No.11275169, No.11075167, No.11105153 and No.11505132. It is also supported in part by the DFG and the NSFC (No.11261130311) through funds provided to the Sino-Germen CRC 110 ``Symmetries and the Emergence of Structure in QCD''. Ning~Li is also supported in part by the Scientific Research Program Funded by Shanxi Provincial Education Department under the grant No.15JK1348.


\end{document}